\documentclass[aps,prd,preprintnumbers,groupedaddress,nofootinbib,amssymb,eqsecnum,notitlepage]{revtex4-1}
\usepackage{here}
\usepackage{graphicx}
\usepackage{amsmath,amsthm,amssymb}
\usepackage{bm}
\usepackage{color}

\usepackage{amsfonts}
\usepackage{dcolumn}
\usepackage{hyperref}
\allowdisplaybreaks[1]
\usepackage{ulem}

\begin{document}
\newcommand{\newc}{\newcommand}

\newcommand{\rk}{\textcolor{red}}
\newc{\ben}{\begin{eqnarray}}
\newc{\een}{\end{eqnarray}}
\newc{\be}{\begin{equation}}
\newc{\ee}{\end{equation}}
\newc{\ba}{\begin{eqnarray}}
\newc{\ea}{\end{eqnarray}}
\newc{\D}{\partial}
\newc{\rH}{{\rm H}}
\newc{\rd}{{\rm d}}
\newc{\rN}{{\rm N}}
\newc{\cX}{{\cal X}}
\newc{\C}{{\cal C}}
\newc{\X}{{\cal X}}
\newc{\Y}{{\cal Y}}

\preprint{WUCG-20-06}

\title{Instability of compact stars with a nonminimal scalar-derivative coupling}

\author{Ryotaro Kase$^{1}$ and Shinji Tsujikawa$^{2}$}

\affiliation{
$^1$Department of Physics, Faculty of Science, 
Tokyo University of Science, 1-3, Kagurazaka,
Shinjuku-ku, Tokyo 162-8601, Japan\\
$^2$Department of Physics, Waseda University, 3-4-1 Okubo, Shinjuku, Tokyo 169-8555, Japan}

\begin{abstract}

For a theory in which a scalar field $\phi$ has a nonminimal derivative coupling to 
the Einstein tensor $G_{\mu \nu}$ of the form $\phi\,G_{\mu \nu}\nabla^{\mu}\nabla^{\nu} \phi$, 
it is known that there exists a branch of static and spherically-symmetric relativistic stars 
endowed with a scalar hair in their interiors. 
We study the stability of such hairy solutions with a radial field dependence $\phi(r)$ 
against odd- and even-parity perturbations. We show that, for the star compactness 
${\cal C}$ smaller than $1/3$,  
they are prone to Laplacian instabilities of the even-parity perturbation associated with 
the scalar-field propagation along an angular direction.
Even for ${\cal C}>1/3$, the hairy star solutions are subject to ghost instabilities. 
We also find that even the other branch with a vanishing background field 
derivative is unstable for a positive perfect-fluid pressure, 
due to nonstandard propagation of 
the field perturbation $\delta \phi$ inside the star.
Thus, there are no stable star configurations in derivative coupling theory 
without a standard kinetic term,  
including both relativistic and nonrelativistic compact objects.

\end{abstract}

\date{\today}


\maketitle

%%%%%%%%%%%%%%%%%%%%%%%%%%%%%%%%%%%%%%%%%%
\section{Introduction}
\label{introsec}
%%%%%%%%%%%%%%%%%%%%%%%%%%%%%%%%%%%%%%%%%%

There have been many attempts for the construction of gravitational 
theories beyond General Relativity (GR) \cite{CST,Clifton,Joyce}. 
This is mostly motivated by the firm 
observational evidence of inflation, dark energy, and 
dark matter \cite{Planck}. 
Usually, new degrees of freedom (DOFs) are taken into account 
to address these problems. One of the candidates for such new 
DOFs is a scalar field with associated potential and kinetic energies. 
If the scalar field is present in today's Universe, it can 
affect the configuration of compact objects.
In particular, after the dawn of gravitational-wave 
astronomy \cite{Abbott2016,GW170817}, it is 
a great concern to search for the signature of new DOFs
beyond GR around strong gravitational objects like black holes (BHs)
and neutron stars (NSs) \cite{Berti,Barack}.

For the canonical scalar field minimally coupled to gravity, the asymptotically flat 
and stationary BH solutions are characterized by only three physical 
quantities--mass, electric charge, and angular 
momentum \cite{Israel,Carter,Wheeler,Hawking,Chase,BekenPRL}. 
This ``no-hair'' BH theorem also holds for scalar-tensor theories in which 
the scalar field $\phi$ is coupled to the Ricci scalar $R$ of the form 
$F(\phi)R$ \cite{Hawking72,Beken95,Soti12}. 
The no-hair property of BHs does not persist in theories with nonminimal scalar 
derivative couplings to the Ricci scalar and Einstein 
tensor \cite{Rinaldi,Anabalon,Minami13,Soti1,Soti2,Babi17}. 
The most general scalar-tensor theories with second-order equations of motion 
accommodating such couplings are known as Horndeski 
theories \cite{Horndeski,Ho1,Ho2,Ho3}. 

In shift-symmetric Horndeski theories, there are conditions 
for the absence of static, spherically-symmetric and asymptotically flat 
BH solutions with a nonvanishing radial scalar derivative $\phi'(r)$. 
This is associated with the conservation of the scalar-field 
current $J_{\phi}^{\mu}$ such that $\nabla_{\mu} J_{\phi}^{\mu}=0$ \cite{Hui}, 
where $\nabla_{\mu}$ is the covariant derivative operator. 
The radial current component $J_{\phi}^r$ can be expressed in the form 
$J_{\phi}^r=\phi' g^{rr} F(\phi';g,g',g'')$, where $g^{rr}$ is the $rr$ component 
of metric tensor $g^{\mu \nu}$ and $F$ contains $\phi'$ and derivatives of 
$g^{\mu \nu}$. Provided that the scalar product 
$g_{\mu \nu} J_{\phi}^{\mu}J_{\phi}^{\nu}$ is regular, 
the regularity of $J_{\phi}^r$ on the BH horizon requires that 
$J_{\phi}^r=0$ everywhere.
If $F$ neither vanishes nor contains negative powers of $\phi'$, 
the allowed field profile consistent with $J_{\phi}^r=0$ is the no-hair solution 
with $\phi'(r)=0$. One counter example is a linearly time-dependent scalar
field $\phi(t, r)=qt+\chi(r)$ \cite{Babi14}, in which case $F=0$ from the field 
equation of motion. This leads to hairy BH solutions 
with a static metric \cite{Koba14,Lefteris,Babi16}, 
including the stealth BH solution \cite{Babi14}. 
A negative function $F$ can be realized by a Gauss-Bonnet term 
linearly coupled to $\phi$, in which case the hairy BH 
is also present \cite{Soti1,Soti2}.

For NSs in scalar-tensor theories, the existence of matter can give rise to 
nontrivial static and spherically-symmetric solutions which do not 
have an analogy with BHs. 
For instance, the nonminimal coupling $F(\phi)R$ allows the presence of 
hairy solutions where the mass and radius of NSs are modified from 
those in GR. This is the case for Brans-Dicke theory and $f(R)$ 
gravity \cite{Cooney:2009rr,Arapoglu:2010rz,Orellana:2013gn,Astashenok:2013vza,
Yazadjiev:2014cza,Resco:2016upv,Kase:2019dqc,Dohi:2020bfs}, 
in which the solutions with nonvanishing $\phi(r)$ are present.
For the nonminimal coupling containing the even power 
of $\phi$, there exists nonrotating scalarized 
NS solutions with $\phi(r) \neq 0$ besides the GR branch 
with $\phi(r)=0$ \cite{Damour,Damour2} 
(see Refs.~\cite{Sotani:2012eb,Doneva:2013qva,Pani:2014jra} for rotating solutions).
If one considers the function $F(\phi)=e^{-\beta \phi^2}$ with a negative coupling 
constant $\beta$, the GR branch can trigger a tachyonic instability to 
reach the other nontrivial branch \cite{Harada:1998ge,Novak:1998rk,Silva:2014fca,Freire:2012mg}, 
whose phenomenon is dubbed spontaneous scalarization. In such theories, the 
scalarized NS solutions are stable against odd- and even-parity 
perturbations on the static and spherically-symmetric background \cite{Kase:2020qvz}.

In shift-symmetric Horndeski theories with matter minimally coupled to 
gravity, there is a no-hair theorem for stars \cite{Lehebel:2017fag} 
generalizing the BH case. 
The theorem states that, under the following three conditions, 
the allowed solution is only the trivial branch with $\phi'(r)=0$: 
(i) $\phi$ and $g_{\mu \nu}$ are regular everywhere, static and 
spherically symmetric, 
(ii) spacetime is asymptotically flat with $\phi' \to 0$ as $r \to \infty$, and
(iii) there is a canonical kinetic term $X=-(1/2)g^{\mu \nu}\nabla_{\mu} \phi \nabla_{\nu} \phi$ 
in the action, where the action is analytic around a trivial scalar-field configuration. 
To realize hairy NS solutions, we need to break at least one of these conditions. 

If we break the condition (iii), the shift-symmetric Lagrangian 
${\cal L}=G_{4}(X) R +G_{4,X}(X)\left[ (\square \phi)^{2}
-(\nabla_{\mu}\nabla_{\nu} \phi)
(\nabla^{\mu}\nabla^{\nu} \phi) \right]$ 
with $G_4(X)=1/(16\pi G_{\rm N})+\eta X/2$ and 
$G_{4,X}={\rm d}G_4/{\rm d}X=\eta/2$, 
where $G_{\rm N}$ is the Newton gravitational constant and $\eta$ is a constant of 
the derivative coupling, can give rise to hairy NS 
solutions \cite{Cisterna:2015yla,Cisterna:2016vdx,Maselli:2016gxk}. 
This theory is equivalent to the nonminimal derivative coupling (NDC) 
$G_5(\phi)=-\eta \phi/2$ with the Lagrangian 
${\cal L}=R/(16\pi G_{\rm N})+G_{5}(\phi)G_{\mu \nu} \nabla^{\mu}
\nabla^{\nu} \phi$, which belongs to a subclass of Horndeski theories.
The NS solutions in NDC theory have an interesting 
property that the scalar hair is present only inside the star 
with an external vacuum \cite{Cisterna:2015yla}. 
At the background level the mass-radius relation in NDC theory 
does not significantly differ from that in GR, but  
the quasi-normal mode of odd-parity perturbations 
exhibits notable difference between the two theories \cite{Blazquez}.

In this paper, we study the stability of hairy NS solutions in NDC theory 
with a radial-dependent field profile $\phi(r)$ by considering 
odd- and even-parity perturbations on the static and spherically-symmetric  
background.  We do not consider a time-dependent scalar-field 
configuration like $\phi(t, r)=qt+\chi(r)$, by reflecting the fact that the BH 
solutions with $q \neq 0$ are generally prone to instabilities against odd-parity 
perturbations \cite{Ogawa,Takahashi}.
We deal with baryonic matter inside the star as a 
perfect fluid described by a Schutz-Sorkin action \cite{Sorkin,Brown,DGS}.
 
The hairy NS solutions in NDC theory are stable against odd-parity 
perturbations with a superluminal radial propagation speed. 
However, we show that 
the angular propagation speed squared of scalar-field perturbation 
$\delta \phi$ in the even-parity sector is negative around the surface 
of NSs for ${\cal C}<1/3$, where ${\cal C}$ is the star compactness. 
This leads to Laplacian instabilities for the perturbations with 
large multipoles $l$ in the angular direction. 
Even for some specific EOSs which give the compactness
${\cal C}>1/3$, there is a ghost instability of even-parity perturbations.
Thus, the compact star solutions in NDC theory with $\phi'(r)\neq 0$ 
are always unstable. 
We also show that, as long as the coupling $G_5(\phi)=-\eta \phi/2$ 
is present, the other branch satisfying $\phi'(r)=0$ 
inside the star is prone to Laplacian instabilities.
These properties are mostly related to the nonstandard 
propagation of $\delta \phi$ induced by the 
absence of standard kinetic term $X$ in the action. 

%%%%%%%%%%%%%%%%%%%%%%%%%%%%%%%%%%%%%%%%%%
\section{Hairy relativistic stars with nonminimal derivative coupling}
\label{nonsec}
%%%%%%%%%%%%%%%%%%%%%%%%%%%%%%%%%%%%%%%%%%

We consider the action of NDC theory given by 
\be
{\cal S}=\int {\rm d}^4 x \sqrt{-g}\,\left[ \frac{1}{16\pi G_{\rm N}} R
+G_{5}(\phi)G_{\mu \nu} \nabla^{\mu}\nabla^{\nu} \phi 
\right]+{\cal S}_m\,,
\label{NDCaction}
\ee
where $g$ is the determinant of metric tensor $g_{\mu \nu}$, and 
\be
G_5(\phi)=-\frac{1}{2} \eta \phi\,,
\label{G5}
\ee
where $\eta$ is a constant.  
After the integration by parts, this theory is equivalent to 
the quartic-order nonminimal derivative coupling 
$G_4(X)=1/(16\pi G_{\rm N})+\eta X/2$ in Horndeski theories \cite{Ho2}.
In other words, the action (\ref{NDCaction}) with the quintic coupling 
(\ref{G5}) belongs to a subclass of shift-symmetric
Horndeski theories invariant under the shift $\phi \to \phi+c$.

For the matter sector, we consider a perfect fluid minimally 
coupled to gravity, which is described by 
the Schutz-Sorkin action \cite{Sorkin,Brown,DGS},
\be
{\cal S}_{m} =  -\int {\rm d}^{4}x \left[
\sqrt{-g}\,\rho(n)
+ J^{\mu} (\partial_{\mu} \ell+{\cal A}_i\partial_{\mu}{\cal B}^i)\right]\,.
\label{SM}
\ee
Here, the matter density $\rho$ is a function of the fluid number density $n$, 
which is related to a vector current field $J^{\mu}$ in the action 
(\ref{SM}), as 
\be
n=\sqrt{\frac{g_{\mu \nu}J^{\mu} J^{\nu}}{g}}\,.
\label{defn}
\ee
The fluid four-velocity $u_{\mu}$ is given by 
\be
u_{\mu}=\frac{J_{\mu}}{n\sqrt{-g}}\,,
\label{defu}
\ee
which satisfies $u^{\mu} u_{\mu}=-1$. 
The scalar quantity $\ell$ in Eq.~(\ref{SM}) is a Lagrange multiplier with the 
notation $\partial_{\mu} \ell \equiv \partial \ell/\partial x^{\mu}$, whereas
the spatial vectors ${\cal A}_i$ and ${\cal  B}^i$ (with $i=1,2,3$)
are the Lagrange multiplier and Lagrangian coordinates of the fluid, 
respectively. Both ${\cal A}_i$ and ${\cal  B}^i$ are nondynamical intrinsic 
vector modes, so that they affect the evolution of dynamical fields 
through constraint equations.
Varying the action (\ref{SM}) with respect to $\ell$, 
it follows that 
\be
\partial_{\mu} J^{\mu}=0\,,
\label{Jmucon}
\ee
which corresponds to the current conservation of 
the perfect fluid.

Since $\partial n/\partial J^{\mu}=-u_{\mu}/\sqrt{-g}$, 
the variation of Eq.~(\ref{SM}) with respect to $J^{\mu}$ gives
\be
\partial_{\mu} \ell = 
\rho_{,n} u_{\mu}
-{\cal A}_i\partial_{\mu}{\cal B}^i\,,
\label{ell}
\ee
where $\rho_{,n} \equiv \partial \rho/\partial n$.
Varying the Lagrangian ${\cal L}_m=-[\sqrt{-g}\,\rho(n)
+ J^{\mu} (\partial_{\mu} \ell+{\cal A}_i\partial_{\mu}{\cal B}^i)]$ 
in Eq.~(\ref{SM}) with respect to $g^{\mu \nu}$, we can derive
the matter energy-momentum tensor $T_{\mu \nu}$. 
On using Eq.~(\ref{ell}) and the relation
$\delta n = (n/2) \left( g_{\mu \nu} -u_{\mu} u_{\nu} \right) \delta g^{\mu \nu}$, 
we obtain
\be
T_{\mu \nu} \equiv
-\frac{2}{\sqrt{-g}} \frac{\delta {\cal L}_m}{\delta g^{\mu \nu}}
=\left( \rho+P \right) u_{\mu} u_{\nu}+P g_{\mu \nu}\,,
\label{Tmunu}
\ee
where $P$ is the matter pressure defined by 
\be
P \equiv n \rho_{,n}-\rho\,.
\ee
Thus, the matter action (\ref{SM}) leads to the standard 
form of perfect-fluid energy-momentum tensor (\ref{Tmunu}). 
This obeys the continuity equation,
\be
\nabla^{\mu} T_{\mu \nu}=0\,.
\label{coneq}
\ee
The current conservation (\ref{Jmucon}) is equivalent to 
the equation $u^{\nu}\nabla^{\mu} T_{\mu \nu}=0$ following 
from Eq.~(\ref{coneq}) \cite{Kase:2020qvz,Amendola:2020ldb}.

We consider a static and spherically-symmetric background 
given by the line element, 
\be
\rd s^2=-f(r) \rd t^{2} +h(r)^{-1} \rd r^{2}
+ r^{2} \left(\rd \theta^{2}+\sin^{2}\theta\,\rd\varphi^{2} 
\right)\,,
\label{BGmetric}
\ee
where $f(r)$ and $h(r)$ are functions of $r$. 
On this background the four velocity in the fluid rest frame 
is given by $u_{\mu}=(-f(r)^{1/2},0,0,0)$, so that 
the energy-momentum tensor (\ref{Tmunu}) 
reduces to 
\be
T^{\mu}_{\nu}={\rm diag} \left( -\rho(r),P(r),P(r),P(r) \right)\,.
\ee
The matter continuity Eq.~(\ref{coneq}) yields 
\be
P'+\frac{f'}{2f} \left( \rho+P \right)=0\,,
\label{mattereq}
\ee
where a prime represents the derivative with respect to $r$.
{}From Eq.~(\ref{defu}), the vector field 
$J^{\mu}$ is expressed as
\be
J^{\mu}=\left( \sqrt{-g}\,n(r) f^{-1/2}(r),0,0,0 \right)\,,
\ee
where $n$ depends on $r$ alone, 
and $\sqrt{-g}=f^{1/2}h^{-1/2}r^2\sin{\theta}$.
Since ${\cal A}_i=0$ on the background (\ref{BGmetric}), 
Eq.~(\ref{ell}) gives
\be
\partial_{\mu} \ell=\left( -\rho_{,n}(r) f(r)^{1/2},0,0,0 
\right)\,.
\ee

For the background scalar field, we consider the configuration, 
\be
\phi=\phi(r)\,.
\ee
Varying the action (\ref{NDCaction}) with respect to $f$ and $h$, 
respectively, we obtain
\ba
h' &=& \frac{1-h-4\pi G_{\rm N}[2 \rho r^2
+\eta h \phi' \{4hr \phi''+(h+1)\phi' \}]}
{r(1+12\pi G_{\rm N} \eta h \phi'^2)}\,,
\label{rheq} \\
f' &=& \frac{f}{h} \frac{1-h+4\pi G_{\rm N} 
[2Pr^2-\eta h \phi'^2 (3h-1)]}
{r(1+12\pi G_{\rm N} \eta h \phi'^2)}\,.
\label{rfeq} 
\ea
Variation of (\ref{NDCaction}) with respect to $\phi$ leads to 
the scalar-field equation $J_\phi'=0$, where 
\be
J_{\phi}=\eta \sqrt{\frac{h}{f}} \phi' 
\left[ f(1-h)-rf' h \right]\,.
\ee
The conservation of $J_{\phi}$ arises from the fact that 
NDC theory with the coupling (\ref{G5})
belongs to a subclass of shift-symmetric Horndeski theories. 
Thus, the equation for $\phi$ reduces to 
$J_{\phi}=C$, where $C$ is an integration constant. 
To satisfy the boundary conditions $f \to 1$, $rf' \to 0$,  
$h \to 1$, and $\phi' \to 0$ at spatial infinity ($r \to \infty$), 
we require that $C=0$. Then, we obtain
\be
\eta \sqrt{\frac{h}{f}} \phi' 
\left[ f(1-h)-rf' h \right]=0\,, 
\label{fhre}
\ee
which means that there are two branches of solutions.
The first one is the trivial branch with a vanishing field derivative, i.e., 
$\phi'(r)=0$ at any distance $r$. 
The second one is the nontrivial branch satisfying 
\be
f'=\frac{f(1-h)}{hr}\,.
\label{feq}
\ee
This is different from the corresponding equation 
$f'=f(1-h)/(hr)+8\pi G_{\rm N}P rf/h$ in GR. 
In other words, the metric component $f$ does not 
feel the matter pressure even inside a star.
The main problem to be addressed in this paper is 
to elucidate whether the branch (\ref{feq}) is stable or not against 
perturbations on the background (\ref{BGmetric}). 
We also discuss the stability of the other branch 
$\phi'=0$ at the end.

Substituting Eq.~(\ref{feq}) into Eq.~(\ref{rfeq}), it follows that 
\be
\eta h \phi'^2=Pr^2\,.
\label{etare}
\ee
For this branch, the field derivative $\phi'$ is related to 
the fluid pressure $P$.
Provided $P>0$, the coupling constant $\eta$ is 
in the range, 
\be
\eta>0\,.
\label{eta}
\ee
We define the star radius $r_s$ at which the pressure 
vanishes, i.e., $P(r_s)=0$. 
Outside the star ($r>r_s$), the field derivative $\phi'$ 
is 0 from Eq.~(\ref{etare}).

Taking the $r$ derivative of Eq.~(\ref{etare}) on account of Eq.~(\ref{mattereq}), 
both $\phi''$ and $\phi'$ can be expressed in terms of 
$\rho$ and $P$. Then, Eq.~(\ref{rheq}) reduces to 
\be
h'=\frac{1-h-4\pi G_{\rm N}r^2[(1+h)\rho+6h P]}
{r(1+4\pi G_{\rm N}r^2 P)}\,.
\label{heq}
\ee
Inside the star, Eq.~(\ref{heq}) differs from the corresponding GR equation 
$h'=(1-h-8\pi G_{\rm N}r^2 \rho)/r$. 
Outside the star we have $\rho=P=0$, so 
Eqs.~(\ref{feq}) and (\ref{heq}) 
are equivalent to the differential equations of $f$ and $h$ in GR respectively.
This means that the effect of NDC on 
the background spacetime appears only in the star interior. 
We define the mass function $M(r)$, as 
\be
h(r)=1-\frac{2G_{\rm N}M(r)}{r}\,,
\label{hr}
\ee
as well as the mass of star, $M_s \equiv M(r_s)$. 
Since $\phi'(r)=0$ outside the star, we have $M(r)=M_s$ 
even at spatial infinity, i.e., $M_s$ corresponds to the ADM mass. 
For $r>r_s$, the metric components are given by 
$f(r)=h(r)=1-2G_{\rm N} M_s/r$.

Around the center of star, we impose the regular 
boundary conditions 
$f(0)=f_c$, $h(0)=1$, $\rho(0)=\rho_c$, $P(0)=P_c$ and 
$f'(0)=h'(0)=\rho'(0)=P'(0)=0$. 
For the hairy branch satisfying Eqs.~(\ref{feq})-(\ref{heq}) 
with Eq.~(\ref{mattereq}), the background solutions expanded 
at $r=0$ are given by 
\ba
f(r) &=& f_c+\frac{4}{3} \pi G_{\rm N} f_c \left( \rho_c+3P_c 
\right)r^2+{\cal O}(r^4)\,,
\label{fex}\\
h(r) &=& 1-\frac{8}{3} \pi G_{\rm N} \left( \rho_c+3P_c \right) 
r^2+{\cal O}(r^4)\,,\\
P(r) &=& P_c-\frac{2}{3} \pi G_{\rm N}   \left( \rho_c+3P_c 
\right) \left( \rho_c+P_c \right)r^2+{\cal O}(r^4)\,,\label{Pex}\\
\phi'^2(r) &=& \frac{P_c}{\eta} r^2+{\cal O}(r^4)\,.
\label{phiex}
\ea
The scalar field satisfies the regular boundary condition 
$\phi'(0)=0$. Note that we can set $f_c=1$ 
by virtue of the time-rescaling invariance of Eq.~(\ref{BGmetric}).
In GR the expansion of $h$ is given by 
$h(r)=1-8\pi G_{\rm N} \rho_c r^2/3+{\cal O}(r^4)$, 
while the forms of $f(r)$ and $P(r)$ are the same as 
Eqs.~(\ref{fex}) and (\ref{Pex}) respectively. 

\begin{figure}[h]
\begin{center}
\includegraphics[height=3.2in,width=3.1in]{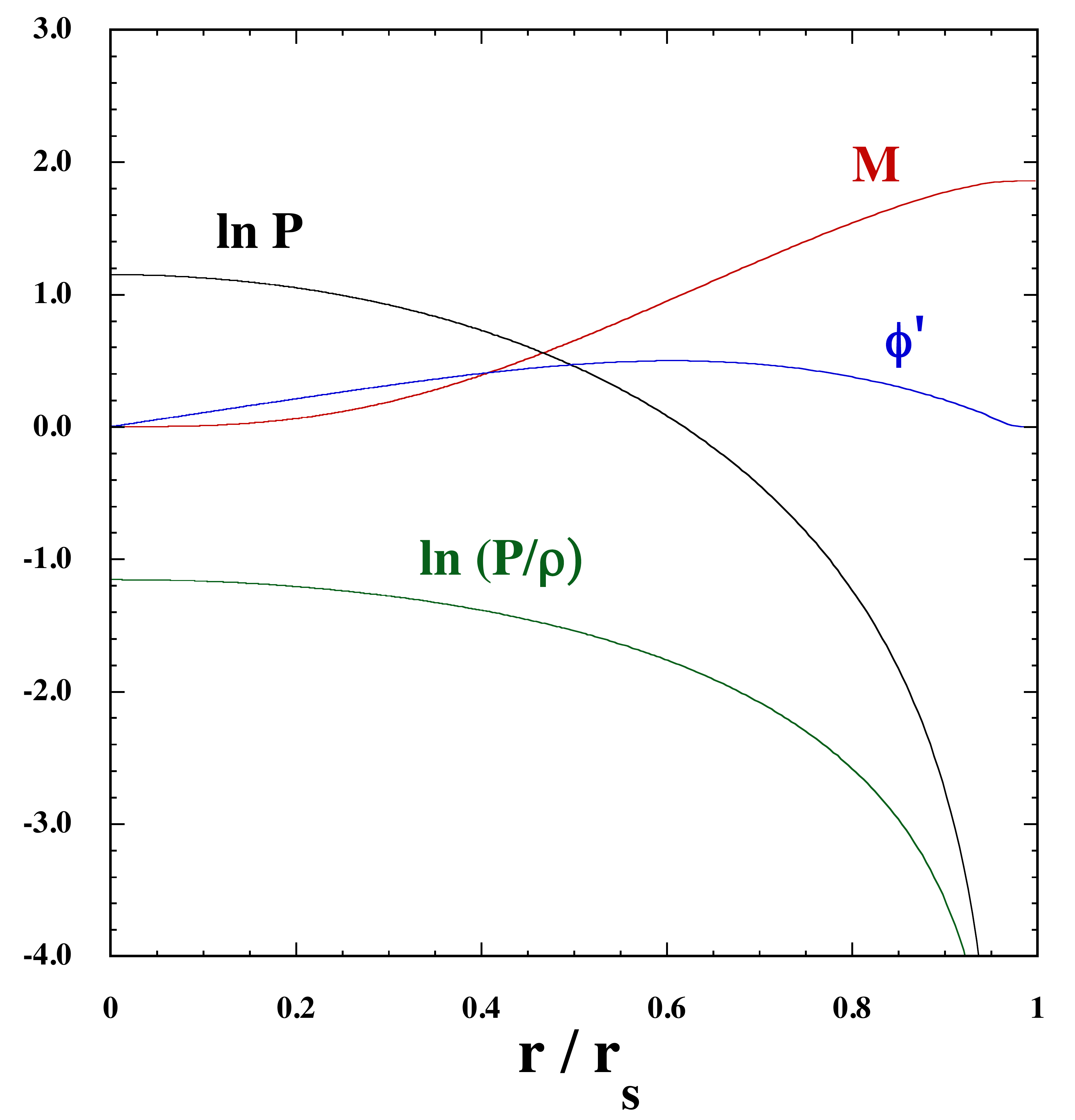}
\includegraphics[height=3.3in,width=3.5in]{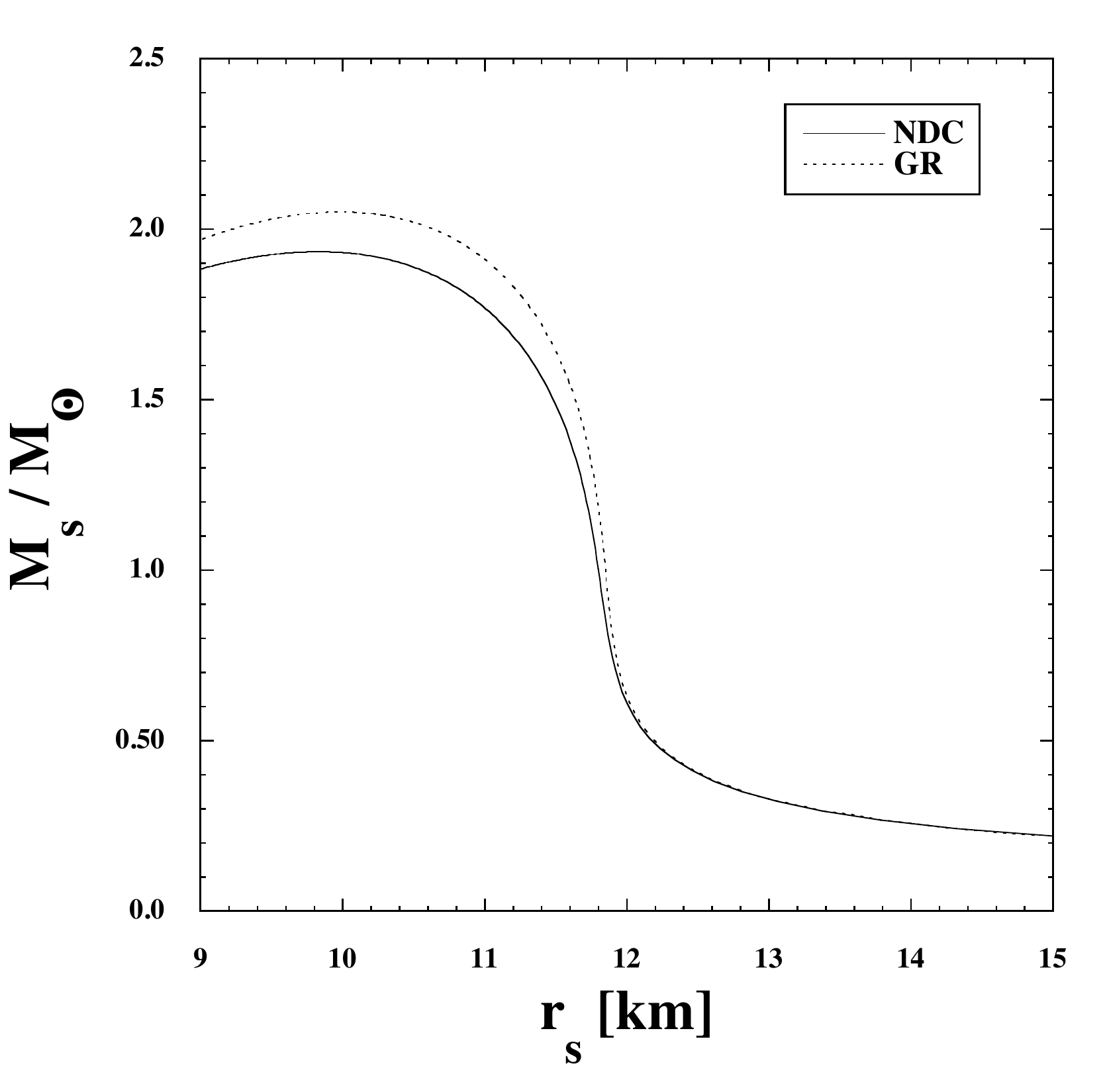}
\end{center}\vspace{-.5cm}
\caption{\label{fig1} 
(Left) $M$, $\phi'$, $\ln P$, and $\ln (P/\rho)$ versus $r/r_s$ 
in NDC theory with the coupling  $\eta=r_s^2$.
We choose the SLy EOS with the central density $\rho_c=10\rho_0$, 
where $\rho_0=1.6749 \times 10^{14}$ g/cm$^3$.
The quantities $M$, $\phi'$, and $P$  are 
normalized by $M_{\odot}$, $(r_s \sqrt{8\pi G_{\rm N}})^{-1}$, 
and $\rho_0$, respectively.
(Right) Mass-radius relations for the SLy EOS in 
NDC theory (solid) and in GR (dashed).
}
\end{figure}

For a given equation of state (EOS) $P=P(\rho)$, the values of 
$f(r)$, $h(r)$, $P(r)$, $\rho(r)$ and $\phi'(r)$ inside the star are known 
by integrating Eqs.~(\ref{mattereq}), (\ref{feq}), and (\ref{heq}) 
with Eq.~(\ref{etare}). 
In Eq.~(\ref{etare}), we choose the branch with $\phi'(r)>0$ 
without loss of generality. 
As an example, we consider the SLy EOS of NSs, whose analytic 
representation is given in Ref.~\cite{Haensel:2004nu}.
In the left panel of Fig.~\ref{fig1}, we plot the mass function $M$, 
field derivative $\phi'$, pressure $P$, ratio $P/\rho$ 
as a function of $r$ for the central density $\rho_c=10\rho_0$
and the coupling $\eta=r_s^2$, where 
$\rho_0=1.6749 \times 10^{14}$ g/cm$^3$.
Since $\eta$ does not appear in the differential 
Eqs.~(\ref{mattereq}), (\ref{feq}), and (\ref{heq}), the star 
configuration is independent of the coupling strength. 
As we see in Eq.~(\ref{etare}), 
for smaller $\eta$, $\phi'$ gets larger.
Up to the star surface the function $M(r)$ continuously grows to 
the ADM mass $M_s$, whereas the pressure $P(r)$ monotonically 
decreases toward 0. In the left panel of Fig.~\ref{fig1}, the 
perfect fluid is in a relativistic region with $P/\rho \simeq 0.315$ 
around the center of star. 
We observe that, as $r$ approaches $r_s$, the ratio $P/\rho$ decreases. 
Around the surface of star, the EOS is in a 
nonrelativistic region characterized by $P/\rho \ll 1$. 
{}From Eq.~(\ref{phiex}), the field derivative grows as 
$\phi' \propto r$ around $r=0$.
Since $\phi'(r_s)=0$ from Eq.~(\ref{etare}), there is a point inside 
the star at which $\phi'(r)$ reaches a maximum. 
This can be confirmed in the numerical simulation of Fig.~\ref{fig1}.

%%%%%%%%%%%%%%%%%%%%%%%%%%%%%%%%
%%%%%%%%%%%%%%%%%%%%%%%%%%%%%%%%

In the right panel of Fig.~\ref{fig1}, we plot $M_s$ (normalized by the 
solar mass $M_{\odot}$) versus $r_s$ for the SLy EOS in NDC theory 
with the branch satisfying (\ref{feq}). 
For the increasing central density $\rho_c$, the radius $r_s$ tends to decrease, 
together with the growth of $M_s$ up to the density  
$\rho_c=14.4 \rho_0$. After $M_s$ reaches the maximum 
value $1.93M_{\odot}$, it starts to decrease for 
$\rho_c>14.4 \rho_0$.
In comparison to the case of GR (plotted as a dashed line), 
the nonvanishing scalar field in NDC theory
works to reduce both $r_s$ and $M_s$. 
Independent of the coupling constant $\eta~(>0)$,  
the mass and radius in NDC theory are uniquely fixed 
for a given EOS. The mass-radius relation is not significantly 
different from that in GR, but the relativistic star is endowed 
with a scalar hair inside the body. 
The difference between NDC theory and GR manifests itself 
for the propagation of perturbations, as 
we will study in the subsequent sections.

%%%%%%%%%%%%%%%%%%%%%%%%%%%%%%%%%%%%%%%%%%
\section{Odd-parity perturbations}
\label{oddsec}
%%%%%%%%%%%%%%%%%%%%%%%%%%%%%%%%%%%%%%%%%%

On the static and  spherically-symmetric background (\ref{BGmetric}), 
the metric perturbations $h_{\mu \nu}$ can be decomposed into odd- and even-parity modes 
with respect to a rotation in the two-dimensional 
plane ($\theta, \varphi$) \cite{Regge:1957td,Zerilli:1970se,Moncrief}. 
In this section, we consider the propagation of odd-parity perturbations
for NDC theory given by the action (\ref{NDCaction}). 
We express the perturbations in terms of the expansion of   
spherical harmonics $Y_{lm}(\theta, \varphi)$. 
For the multipoles $l \geq 2$ we choose the so-called Regge-Wheeler 
gauge \cite{Regge:1957td} in which the components $h_{ab}$, 
where $a, b$ is either $\theta$ or $\varphi$, vanish. 
The nonvanishing odd-parity metric perturbation components are 
then given by 
\be
h_{ta}=\sum_{l,m}Q(t,r)E_{ab}\nabla^bY_{lm}(\theta,\varphi)\,,
\qquad
h_{ra}=\sum_{l,m}W(t,r)E_{ab} \nabla^bY_{lm}(\theta,\varphi)\,,
\ee
where $Q$ and $W$ are functions of $t$ and $r$.
The tensor $E_{ab}$ is defined by $E_{ab}=\sqrt{\gamma}\,\varepsilon_{ab}$, where 
$\gamma$ is the determinant of two-dimensional metric 
$\gamma_{ab}$ and $\varepsilon_{ab}$ is the anti-symmetric symbol 
with $\varepsilon_{\theta\varphi}=1$. 

For the odd-parity sector of the perfect fluid, the components of 
$J_{\mu}$ in the action (\ref{SM}) can be 
expressed as \cite{Kase:2020qvz}
\be
J_t=-n(r) \sqrt{f(r)} \sqrt{-\bar{g}}\,,\qquad
J_{r}=0\,,\qquad 
J_{a}=\sum_{l,m} \sqrt{-\bar{g}}\,\delta j(t,r)E_{ab}
\nabla^bY_{lm}(\theta,\varphi)\,,
\label{Jodd}
\ee
where $\sqrt{-\bar{g}}=\sqrt{f(r)/h(r)}\,r^2 \sin \theta$ 
is the background value, and the components 
$J_a$ have a perturbation $\delta j(t,r)$. 
The intrinsic vectors ${\cal A}_i$ and ${\cal B}^i$ 
can be chosen in the forms, 
\be
{\cal A}_i=\delta {\cal A}_i\,,\qquad 
{\cal B}^i=x^i+\delta {\cal B}^i\,,
\label{ABi}
\ee
with the odd-parity perturbations, 
\ba
&&
\delta {\cal A}_r=0\,,\qquad 
\delta {\cal A}_a=\sum_{l,m}\delta {\cal A}(t,r)
E_{ab} \nabla^bY_{lm}(\theta,\varphi)\,,\\
&&
\delta {\cal B}^r=0\,,\qquad 
\delta {\cal B}^a=\sum_{l,m}\delta {\cal B}(t,r)
{E^{a}}_{b} \nabla^bY_{lm}(\theta,\varphi)\,.
\label{delB}
\ea
The Lagrange multiplier $\ell$ in Eq.~(\ref{ell}) is not affected 
by the odd-parity perturbation, so that its explicit form 
without containing the even-parity perturbation is given by 
\be
\ell=-\rho_{,n} (r) \sqrt{f(r)}\,t\,.
\label{lex}
\ee
{}From Eq.~(\ref{defn}), the perturbation of 
fluid number density $n$ is expressed as
\be
\delta n=\frac{n}{2r^2}\left(hW^2-\frac{2}{f}Q^2
+\frac{2}{n\sqrt{f}}Q\delta j-\frac{\delta j^2}{n^2}\right)
\left[ (\partial_{\theta}Y_{lm})^2+\frac{(\partial_{\varphi}
Y_{lm})^2}{\sin^2\theta} \right]
+{\cal O}(\varepsilon^4)\,, 
\label{deln}
\ee
with $\rho=\rho(r)+\rho_{,n}\delta n+{\cal O}(\varepsilon^4)$, 
where $\varepsilon^i$ represents the $i$-th order of perturbations. 
Expanding the Schutz-Sorkin action (\ref{SM}) up to the order of 
$\varepsilon^2$, the resulting second-order action contains a term proportional 
to $\delta {\cal A}$. The variation of this term with respect to 
$\delta {\cal A}$ leads to 
\be
\dot{\delta {\cal B}}=\frac{nQ-\sqrt{f}\,\delta j}{nr^2}\,,
\ee
where a dot represents the derivative with respect to $t$.
Substituting this relation into ${\cal S}_m$, the resulting quadratic-order 
action contains the Lagrangian, 
\be
{\cal L}_{\delta j}=\frac{\sqrt{f} (\rho+P) 
[(\partial_{\theta}Y_{lm})^2\sin^2 \theta +
(\partial_{\varphi}Y_{lm})^2]}{2n^2 \sqrt{h} \sin \theta} \delta j^2\,.
\ee
Varying ${\cal L}_{\delta j}$ with respect to $\delta j$, we obtain
\be
\delta j=0\,,
\label{delj}
\ee
which means that the perturbations of $J_{\mu}$ in the odd-parity 
sector vanish.
On using Eq.~(\ref{delj}), the matter perturbations arising from 
the Schutz-Sorkin action are integrated out from 
the total action (\ref{NDCaction}).
This is analogous to the case of cosmological perturbations in the 
presence of a perfect-fluid action (see, e.g., Ref.~\cite{Kase:2018aps}). 
On the cosmological background in scalar-tensor theories, the matter 
perturbation arising from the Schutz-Sorkin action does not contribute 
to tensor perturbations, but it only modifies 
the dynamics of scalar perturbations. 
On the static and spherically-symmetric background the scalar perturbation 
corresponds to the even-parity mode, so the odd-parity perturbation 
$\delta j$ in the Schutz-Sorkin action vanishes. 

The scalar field $\phi$ does not have the odd-parity perturbation. 
We integrate the total second-order action with respect to 
$\theta$ by setting $m=0$ without loss of generality. 
After the integration by parts with respect to $t$ and $r$,
the quadratic-order action of odd-parity perturbations reduces to
\be
{\cal S}^{(2)}_{\rm odd}=\sum_{l}
\int \rd t \rd r \left[ \frac{L}{4} \sqrt{\frac{h}{f}}{\cal H}
\left( \dot{W}-Q'+\frac{2Q}{r} \right)^2
-\frac{L(L-2)\sqrt{fh}}{4r^2}{\cal G}W^2
+\frac{L(L-2)}{4\sqrt{fh}\,r^2}{\cal F}Q^2 \right]\,,
\label{Sodd}
\ee
where 
\be
L=l(l+1)\,,
\ee
and 
\be
{\cal H} = {\cal G} = \frac{1}{8\pi G_{\rm N}}+\frac{1}{2} 
\eta h \phi'^2 \,,\qquad
{\cal F} = \frac{1}{8\pi G_{\rm N}}
-\frac{1}{2} \eta h \phi'^2  \,.
\label{Fexp}
\ee
For the derivation of Eq.~(\ref{Sodd}) we did not choose the branch of 
either $\phi'(r)=0$ or $\phi'(r) \neq 0$, so the result is valid for both cases.
The second-order action (\ref{Sodd}) does not contain any matter perturbations 
arising from the Schutz-Sorkin action (\ref{SM}).
The existence of a perfect fluid inside the star affects the dynamics of 
odd-parity perturbations only through the background metric components 
$f$ and $h$. 
Since the matter perturbations are fully integrated out from the odd-parity action, 
the result (\ref{Sodd}) coincides with the second-order action of 
Ref.~\cite{Kobayashi:2012kh} derived in the context of BH perturbations 
for full Horndeski theories.

Although there are two metric perturbations $W$ and $Q$ in 
Eq.~(\ref{Sodd}), the system can be described by a single 
dynamical perturbation \cite{DeFelice:2011ka,Kase:2014baa}, 
\be
\chi= \dot{W}-Q'+\frac{2Q}{r}\,.
\label{chiex}
\ee
To see this property,  we express the action (\ref{Sodd}) in the form, 
\be
{\cal S}^{(2)}_{\rm odd}=\sum_{l} \int 
\rd t \rd r \left\{ \frac{L}{4} \sqrt{\frac{h}{f}} {\cal H}
\left[ 2\chi \left( \dot{W}-Q'+\frac{2Q}{r} \right)-\chi^2 
\right]
-\frac{L(L-2)\sqrt{fh}}{4r^2}{\cal G}W^2
+\frac{L(L-2)}{4\sqrt{fh}\,r^2}{\cal F}Q^2 \right\}\,.
\label{Sodd2}
\ee
Varying Eq.~(\ref{Sodd2}) with respect to 
$W$ and $Q$, it follows that both 
$W$ and $Q$ can be expressed in terms of $\chi$ and its 
$t$ and $r$ derivatives.
Substituting these relations into Eq.~(\ref{Sodd2}) and 
integrating it by parts, the second-order action 
for $l \geq 2$ reduces to 
\be
{\cal S}^{(2)}_{\rm odd}=\sum_{l} \int 
\rd t \rd r \left( C_1 \dot{\chi}^2+C_2 \chi'^2
+C_3 \chi^2 \right)\,,
\label{Sodd3}
\ee
where
\be
C_1 = \frac{L\sqrt{h}\,r^2 {\cal H}^2}{4f^{3/2}(L-2){\cal G}}\,,\qquad
C_2 = -\frac{Lh^{3/2}r^2 {\cal H}^2}{4\sqrt{f} (L-2){\cal F}}\,,\qquad
C_3 = -\frac{L\sqrt{h} {\cal H}}{8 (L-2)f^{5/2}{\cal F}^2}\tilde{C_3}\,,
\ee
and
\ba
\tilde{C_3}  &=& 2(L-2) f^2 {\cal F}^2+f^2 r {\cal F}' \left[ (h'{\cal H}+2h{\cal H}')r+4h{\cal H} 
\right]-4f^2 {\cal F} \left[  (h'{\cal H}+h{\cal H}')r-h{\cal H} \right]
-f'^2h r^2 {\cal F} {\cal H}\nonumber \\
& &
-f^2 r^2{\cal F} \left( 2h {\cal H}''+3h'{\cal H}'+h'' {\cal H} \right) 
+fr^2 \left[ {\cal F} \left( f'' h {\cal H}+f' h'{\cal H}+f'h {\cal H}' \right)
-f' h {\cal F}' {\cal H} \right]\,.
\ea
In the following, we focus on the hairy branch satisfying the condition (\ref{etare})
by the end of this section. 
{}From Eq.~(\ref{Sodd3}) the ghost is absent under the condition 
$C_1>0$, which translates to
\be
{\cal G}=\frac{1}{8\pi G_{\rm N}}
+\frac{1}{2} Pr^2>0\,.
\label{con1}
\ee
For $P>0$, this condition is automatically satisfied.

The dispersion relation in the radial direction follows by substituting 
the solution $\chi=e^{i(\omega t-kr)}$ into Eq.~(\ref{Sodd3}) and taking 
the limits of large $\omega$ and $k$, such that 
$\omega^2 C_1+k^2 C_2=0$. 
The associated propagation speed squared in proper time 
is given by $c_r^2=(\omega^2/k^2)/(fh)=-(C_2/C_1)/(fh)$, i.e., 
\be
c_r^2=\frac{{\cal G}}{{\cal F}}
=\frac{1+4\pi G_{\rm N}P r^2}{1-4\pi G_{\rm N}P r^2}\,.
\label{crodd}
\ee
The Laplacian instability in the radial direction is 
absent for 
\be
4\pi G_{\rm N}P r^2<1\,.
\label{lapcon}
\ee
Provided that $P \lesssim \rho$, the quantity $4\pi G_{\rm N}P r^2$ 
is at most of the order of the star compactness ${\cal C}=G_{\rm N}M_s/r_s$.
Hence the condition (\ref{lapcon}) is well satisfied for NSs with 
${\cal C} \lesssim 0.3$. 
Indeed, in the numerical simulation of Fig.~\ref{fig1}, 
we confirmed that the condition (\ref{lapcon}) 
holds inside the star. 
{}From Eq.~(\ref{crodd}), the radial propagation of odd-parity perturbations 
is superluminal ($c_r^2>1$) for $0<r<r_s$. Around $r=0$ the pressure $P$ 
is given by Eq.~(\ref{Pex}), so that $c_r^2=1$ at $r=0$. 
For increasing $r$ inside the star, $c_r^2$ first increases toward 
the superluminal region and then it starts to decrease at some 
radius to approach the value $c_r^2=1$ as $r \to r_s$.

The speed of propagation $c_{\Omega}$ in the angular direction 
follows by plugging the solution of the form 
$\chi=e^{i(\omega t-l \theta)}$ into Eq.~(\ref{Sodd3}) 
and taking the limit $l \to \infty$. In this case the dispersion relation yields 
$\omega^2 C_1+C_3=0$, 
with $C_1 \simeq \sqrt{h} r^2 {\cal H}^2/(4f^{3/2}{\cal G})$ and $
C_3 \simeq -l^2 \sqrt{h} {\cal H}/(4\sqrt{f})$. 
In proper time, the propagation speed squared is given by 
$c_{\Omega}^2=(\omega^2 r^2/l^2)/f=-C_3 r^2/(C_1 l^2 f)$, 
so that 
\be
c_{\Omega}^2=\frac{{\cal G}}{{\cal H}}=1\,.
\label{coodd}
\ee
Hence the Laplacian instability is absent along the angular direction, 
with $c_{\Omega}$ equivalent to that of light.

As in Ref.~\cite{Kase:2020qvz}, the perturbation (\ref{chiex}) for $l=1$ 
does not propagate as a dynamical degree of freedom, so 
there are no additional stability conditions arising from the dipole mode.

%%%%%%%%%%%%%%%%%%%%%%%%%%%%%%%%%%%%%%%%%%
\section{Even-parity perturbations}
\label{evensec}
%%%%%%%%%%%%%%%%%%%%%%%%%%%%%%%%%%%%%%%%%%

Let us proceed to the stability analysis against even-parity perturbations
for the multipoles $l \geq 2$. 
We choose the gauge in which the components $h_{ta}$ and 
$h_{ab}$ of metric perturbations $h_{\mu \nu}$ vanish. 
Then, the nonvanishing components of $h_{\mu \nu}$ are given by 
\ba
&&
h_{tt}=f(r) \sum_{l,m}H_0(t,r)Y_{lm}(\theta,\varphi)\,,\qquad
h_{tr}=h_{rt}=\sum_{l,m}H_1(t,r)Y_{lm}(\theta,\varphi)\,,\nonumber \\
&&
h_{rr}=h(r)^{-1}\,\sum_{l,m}H_2(t,r)Y_{lm}(\theta,\varphi)\,,\qquad
h_{ra}=h_{ar}=\sum_{l,m}\alpha(t,r)\nabla_aY_{lm}(\theta,\varphi)\,,
\label{meteven}
\ea
where $H_0$, $H_1$, $H_2$, $\alpha$ are perturbed quantities.
We also expand the scalar field in the form,
\be
\phi=\phi(r)+\sum_{l,m} \delta \phi (t,r) 
Y_{lm} (\theta,\varphi)\,,
\label{phidec}
\ee
where $\phi(r)$ is the background value, and $\delta \phi$ 
is the scalar perturbation.

For the perfect fluid, the current $J_{\mu}$ in the even-parity 
sector contains three metric perturbations $\delta J_t(t, r)$, 
$\delta J_r(t, r)$, and $\delta J(t, r)$, as 
\be
J_{t}=\bar{J}_t+\sum_{l,m} \sqrt{-\bar{g}}\,\delta J_t (t,r)Y_{lm}(\theta,\varphi)\,,\qquad
J_{r}=\sum_{l,m} \sqrt{-\bar{g}}\,\delta J_r (t,r)Y_{lm}(\theta,\varphi)\,,\qquad
J_{a}=\sum_{l,m} \sqrt{-\bar{g}}\,\delta J(t,r) \nabla_aY_{lm}(\theta,\varphi)\,,
\label{Jeven}
\ee
where $\bar{J}_t=-n(r) (f(r)/\sqrt{h(r)}) r^2 \sin \theta$.
The intrinsic spatial vector fields ${\cal A}_i$ and ${\cal B}_i$ 
are chosen as Eq.~(\ref{ABi}) with the perturbed components,
\ba
& &
\delta {\cal A}_r=\sum_{l,m}\delta {\cal A}_1(t,r)Y_{lm}(\theta,\varphi)\,,\qquad 
\delta {\cal A}_a=\sum_{l,m}\delta {\cal A}_2(t,r)\nabla_aY_{lm}(\theta,\varphi)\,,
\label{delAa}\\
&&
\delta {\cal B}_r=\sum_{l,m}\delta {\cal B}_1(t,r)Y_{lm}(\theta,\varphi)\,,\qquad 
\delta {\cal B}_a=\sum_{l,m}\delta {\cal B}_2(t,r)\nabla_aY_{lm}(\theta,\varphi)\,.
\label{delBa}
\ea
The density $\rho$ in the action (\ref{SM}) is expanded as
\be
\rho=\rho(r)+\rho_{,n}\delta n
+\frac{\rho_{,n}}{2n}c_m^2  \delta n^2+
{\cal O}(\varepsilon^3)\,,
\ee
where $\rho(r)$ is the background value, and $c_m^2$ is 
the matter sound speed squared defined by 
\be
c_m^2=\frac{n \rho_{,nn}}{\rho_{,n}}\,.
\label{cm}
\ee
We define the matter density perturbation $\delta \rho (t,r)$ 
according to 
\be
\delta n=\sum_{l,m} \frac{\delta \rho (t,r)}{\rho_{,n}(r)} 
Y_{lm} (\theta, \varphi)\,.
\label{delnrho}
\ee
The $\theta, \varphi$ components of rotational-free four velocity $u_{\mu}$ 
are related to the velocity potential $v(t,r)$, as
\be
u_a=\sum_{l,m} v(t,r) \nabla_a Y_{lm}(\theta,\varphi)\,.
\label{ua}
\ee
{}From Eq.~(\ref{ell}), we have $\partial_a \ell=\rho_{,n} u_a-\delta {\cal A}_a$
up to first order in even-parity perturbations.
Integrating this relation with respect to $a$ on account of 
Eqs.~(\ref{delAa}) and (\ref{ua}), it follows that 
\be
\ell=-\rho_{,n} (r) \sqrt{f(r)}\,t+\sum_{l,m} \left[ \rho_{,n}(r) v(t,r)-\delta {\cal A}_2 
(t,r) \right] Y_{lm} (\theta, \varphi)\,.
\label{ldef}
\ee

For the expansion of the total action (\ref{NDCaction}), we perform the integral 
with respect to $\theta$ by setting $m=0$ without loss of generality.
Following the same procedure as in Ref.~\cite{Kase:2020qvz}, the nondynamical 
variables $\delta {\cal A}_1$,  $\delta {\cal A}_2$,  $\delta {\cal B}_1$, $\delta {\cal B}_2$, 
$\delta J$, and $\delta J_r$ can be integrated out from the second-order action 
of even-parity perturbations.
The resulting quadratic-order action is given by 
${\cal S}_{\rm even}^{(2)}=\sum_{l} \int \rd t \rd r {\cal L}$, 
with the Lagrangian
\ba
{\cal L}
&=& H_0 \left[ a_1 \delta \phi'' +a_2 \delta \phi' +a_3 H_2'
+L a_4 \alpha'+\left( a_5+L a_6 \right) \delta \phi 
+\left( a_7+L a_8 \right) H_2+L a_9 \alpha +a_{10} \delta \rho \right] \nonumber \\
& &+L b_1 H_1^2+H_1 \left( b_2 \dot{\delta \phi}'+b_3 \dot{\delta \phi}
+b_4 \dot{H}_2+L b_5 \dot{\alpha} \right)
+c_1 \dot{\delta \phi} \dot{H}_2
+H_2 \left[ c_2 \delta \phi'+ (c_3+L c_4) \delta \phi
+L c_5 \alpha+\tilde{c}_5 \dot{v} \right]+c_6 H_2^2 \nonumber \\
& &+L d_1 \dot{\alpha}^2+L \alpha \left( d_2 \delta \phi'
+d_3 \delta \phi \right)+L d_4 \alpha^2
+e_1 \dot{\delta \phi}^2+e_2 \delta \phi'^2
+\left( e_3+L e_4 \right) \delta \phi^2
+L f_1 v^2+f_2 \delta \rho^2+f_3 \delta \rho\,\dot{v}\,,
\label{eaction}
\ea
where the coefficients $a_1$ etc are given in Appendix.
Variations of the Lagrangian (\ref{eaction}) with respect to 
$H_0$, $H_1$, and $v$ lead, respectively, to
\ba
& &
a_1 \delta \phi'' +a_2 \delta \phi' +a_3 H_2'
+L a_4 \alpha'+\left( a_5+L a_6 \right) \delta \phi 
+\left( a_7+L a_8 \right) H_2+L a_9 \alpha +a_{10} \delta \rho=0\,,
\label{econ1}\\
& &
2L b_1 H_1+b_2 \dot{\delta \phi}'+b_3 \dot{\delta \phi}
+b_4 \dot{H}_2+L b_5 \dot{\alpha}=0\,,
\label{econ2}\\
& &
2L f_1v-f_3 \dot{\delta \rho}-\tilde{c}_5 \dot{H}_2=0\,.
\label{econ3}
\ea

Besides the scalar-field perturbation $\delta \phi$, there are dynamical 
perturbations arising from the gravity and perfect-fluid sectors. 
To study the propagation of latter two perturbations, we introduce 
the combinations \cite{Kobayashi:2014wsa,Kase:2020qvz}, 
\ba
\psi &\equiv& a_3 H_2+L a_4 \alpha+a_1 \delta \phi'\,,
\label{psi}\\
\delta \rho_m &\equiv& \delta \rho+\frac{2\sqrt{f}\,hr^3 f_1}
{f_3[ha_3(2f-f' r)+Lfr a_4]}\psi'
-\frac{h f_1 r^2[a_1(2f-f' r)+f^2 rb_3]}
{\sqrt{f} f_3[ha_3(2f-f' r)+Lfr a_4]}\delta \phi'\,.
\label{delrhom}
\ea
Taking the $r$ derivative of Eq.~(\ref{psi}), the derivatives 
$a_1 \delta \phi''$, $a_3 H_2'$, and $L a_4 \alpha'$ can be 
simultaneously eliminated from Eq.~(\ref{econ1}), so that 
$\alpha$ can be expressed in terms of 
$\psi$, $\delta \phi$, and $\delta \rho_m$ and their first radial derivatives. 
We also differentiate Eq.~(\ref{psi}) with respect to $t$ and eliminate 
$\dot{H}_2$ in Eqs.~(\ref{econ2}) and (\ref{econ3}) to solve for 
$H_1$ and $v$, respectively. 
Substituting these relations into Eq.~(\ref{eaction}), 
the second-order action of even-parity perturbations 
reduces to the form, 
\be
{\cal S}_{\rm even}^{(2)}=\sum_{l} \int \rd t \rd r
\left( 
\dot{\vec{\mathcal{X}}}^{t}{\bm K}\dot{\vec{\mathcal{X}}}
+\vec{\mathcal{X}}^{'t}{\bm G}\vec{\mathcal{X}}^{'}
+\vec{\mathcal{X}}^{t}{\bm Q}\vec{\mathcal{X}}^{'}
+\vec{\mathcal{X}}^{t}{\bm M} \vec{\mathcal{X}}
\right)\,,
\label{Ss2}
\ee
with the dynamical perturbations,  
\be
\vec{\mathcal{X}}^{t}=\left( \delta \rho_m, \psi, \delta \phi \right) \,.
\label{calX}
\ee
The components of $3 \times 3$ matrices ${\bm K}$, ${\bm G}$, ${\bm Q}$, ${\bm M}$, 
which we denote $K_{ij}$ etc, 
determine the stability of stars against even-mode perturbations. 

In the following, we will consider the branch in NDC theory 
obeying Eqs.~(\ref{feq})-(\ref{heq}). 
At the end of this section, we also study the stability of the 
other branch $\phi'(r)=0$. 

First of all, there are no ghosts under the following three conditions,
\ba
& &
K_{11}>0\,,\label{go1}\\
& &
K_{11}K_{22}-K_{12}K_{21}>0\,,\label{go2}\\
& &
{\rm det}\,{\bm K}>0\,.\label{go3}
\ea
The inequality (\ref{go1}) is satisfied for
\be
\rho+P>0\,,
\label{go1a}
\ee
which holds for standard baryonic matter.
Under the condition (\ref{go1a}), 
the inequality (\ref{go2}) translates to
\ba
& &
h(L-2)\left( 1+20\pi G_{\rm N} P r^2 \right)+4\pi G_{\rm N} L r^2 
\left( 1+4\pi G_{\rm N} P r^2 \right) \left[ \rho (1+
12\pi G_{\rm N} hP r^2 )+P \right] \nonumber \\
& &+32P^2 \pi^2 G_{\rm N}^2 (11L-3)h  r^4
+1152 P^3  \pi^3  G_{\rm N}^3 (L+1) h r^6>0\,,
\ea
which is trivially satisfied for $l \geq 2$. 
As long as the conditions (\ref{eta}) and 
(\ref{go1a}) are satisfied, the third condition (\ref{go3}) amounts to 
\ba
{\cal K} &\equiv&
\rho (3h-1) \left[ (L-2)h+4\pi G_{\rm N} L \rho r^2 
\left( 1+4\pi G_{\rm N} P r^2 \right) \right]
+P \left[ h(19h-1)(L-2)+8\pi G_{\rm N} \rho r^2 
\left\{ L (13h-1)-4h \right\} \right] \nonumber \\
& &+ 4P^2 \pi G_{\rm N} r^2 \left[ 8(L-2)h^2+
L(23h-1)-8h-4\pi G_{\rm N} \rho r^2 \left\{ 
3(L-2)h^2-L(27h-2)-6h \right\} \right] \nonumber \\
& &-16 P^3 \pi^2 G_{\rm N}^2 r^4 
\left[ 27(L-2)h^2-L(24h-1)-6h \right]>0\,.
\label{go3a}
\ea
On using the expansions (\ref{fex})-(\ref{Pex}) around $r=0$, 
the dominant contribution to the right-hand side of Eq.~(\ref{go3a}) is given by 
$2(L-2)(\rho_c+9P_c)$, so the inequality (\ref{go3a}) holds 
around the center of star. For increasing $r$, the effect of pressure $P$ tends 
to be negligible relative to the density $\rho$, see the left panel of Fig.~\ref{fig1}. 
Around $r=r_s$, the dominant contribution to ${\cal K}$ in Eq.~(\ref{go3a}) 
is the first term containing $\rho$, i.e., 
\be
{\cal K}_s \equiv \rho (3h-1) 
\left[ (L-2)h+4\pi G_{\rm N} L \rho r^2 \right]\,.
\label{Ks}
\ee
The positivity of ${\cal K}_s$ requires that  
\be
h>\frac{1}{3}\,.
\label{hcon}
\ee
For the SLy EOS used in the numerical simulation of Fig.~\ref{fig1}, 
we confirmed that the no-ghost condition (\ref{go3a}) holds 
inside the star irrespective of the central density $\rho_c$, 
with $h>1/3$ around $r=r_s$.
This is also the case for other NS EOSs like 
FPS \cite{Haensel:2004nu} and 
BSk19, 20, 21 \cite{Potekhin:2013qqa}.

Let us proceed to the discussion for the propagation of even-parity 
perturbations along the radial direction. In the small-scale limit, 
the speed $c_r$ in proper time can be derived by solving
\be
{\rm det} \left| fh c_r^2 {\bm K}+{\bm G} \right|=0\,.
\label{crso}
\ee
{}From Eq.~(\ref{ua}) the matter velocity potential $v$ is related to only 
the $\theta$ and $\varphi$ components of $u_{\mu}$. 
Hence there is no propagation 
of matter perturbations $\delta \rho_m$ in the radial direction. 
This property manifests itself as the vanishing matrix components 
$G_{11}$, $G_{12}$, and $G_{13}$ of ${\bm G}$, so that 
the first solution to Eq.~(\ref{crso}) is given by 
\be
c_{r1}^2=0\,. 
\label{crm}
\ee
The other two propagation speed squares are
\ba
c_{r\pm}^2=\frac{\alpha_1\pm \sqrt{\alpha_1^2-\alpha_2 {\cal K}}}
{{\cal K}}\,,
\ea
where ${\cal K}$ is defined by Eq.~(\ref{go3a}), and 
\ba
\alpha_1 &=& \frac{\rho}{2} \left( 3h-1 \right) \left[ (L-2)h
+4\pi G_{\rm N} L c_m^2 \rho r^2 (1+4\pi G_{\rm N}Pr^2)\right] 
+\frac{P}{2} [ (L-2) h (19h+3)+\pi G_{\rm N} \rho r^2 \{ 
24 (L-2)h^2 \nonumber \\
& & 
+8 (13c_m^2+1) hL-16 (2 c_m^2+1)h-8L (c_m^2-2) \}]
+P^2 r^2 \pi G_{\rm N} [ r^2 \{ 8 (26c_m^2+3)hL+16(4c_m^2-3)h
\nonumber \\
& & 
-16 L(c_m^2-2) \} \pi G_{\rm N} \rho  
+4h^2(L-2)(23+6\pi G_{\rm N} \rho r^2)+2(23c_m^2+2)h L-8(2c_m^2+1)h-2L(c_m^2-4) ]
\nonumber \\
& & 
+8P^3 r^4 \pi^2 G_{\rm N}^2 \left[ 27(L-2)h^2
+(23c_m^2-1)hL+2(4c_m^2+1)h-L(c_m^2-4) \right]\,,\\
\alpha_2 &=& 4P(1+4\pi G_{\rm N}Pr^2)
\left[ (L-2)h(1+4\pi G_{\rm N}Pr^2)+
4\pi G_{\rm N} c_m^2 (\rho+P) r^2
\{(L-2)h+L\} \right]\,.
\ea
Exploiting the background solutions (\ref{fex})-(\ref{Pex}) 
around $r=0$, it follows that 
\ba
c_{r+}^2 &=& 1+{\cal O}(r^2)\,,
\label{crp0}\\
c_{r-}^2 &=& \frac{2P_c}{\rho_c+9P_c}+{\cal O}(r^2)\,,
\label{crm0}
\ea
whose leading-order terms are both positive.

Around $r=r_s$, the perfect fluid is in the nonrelativistic regime 
characterized by $P/\rho \ll 1$. Then, we have 
\ba
c_{r+}^2 &=&\frac{(L-2)h+4\pi G_{\rm N}L c_m^2 \rho r^2}
{(L-2)h+4\pi G_{\rm N}L \rho r^2}+{\cal O}\left( \frac{P}{\rho} \right)\,,
\label{cpl}\\
c_{r-}^2 &=& \frac{4(L-2)h(1+4\pi G_{\rm N} c_m^2 \rho r^2)
+16\pi G_{\rm N}L c_m^2 \rho r^2}
{(3h-1)[(L-2)h+4\pi G_{\rm N}L c_m^2 \rho r^2]}\frac{P}{\rho}
+{\cal O}\left( \frac{P^2}{\rho^2} \right)\,.
\ea
In this region, $c_{r+}^2$ is subluminal for $0\leq c_m^2\leq1$ 
and approaches 1 as $r \to r_s$. 
We note that $c_{r+}$ corresponds to the propagation speed 
of the gravitational perturbation $\psi$, which smoothly 
joins the external value 1 at $r=r_s$. 
On the other hand, $c_{r-}$ is the speed of the scalar-field 
perturbation $\delta \phi$.
Provided that $h>1/3$, $c_{r-}^2$ is positive around the 
surface of star and approaches $+0$ as $r \to r_s$.

%%%%%%%%%%%%%%%%%%%%%%%%%%%%%%%%
\begin{figure}[h]
\begin{center}
\includegraphics[height=3.3in,width=3.2in]{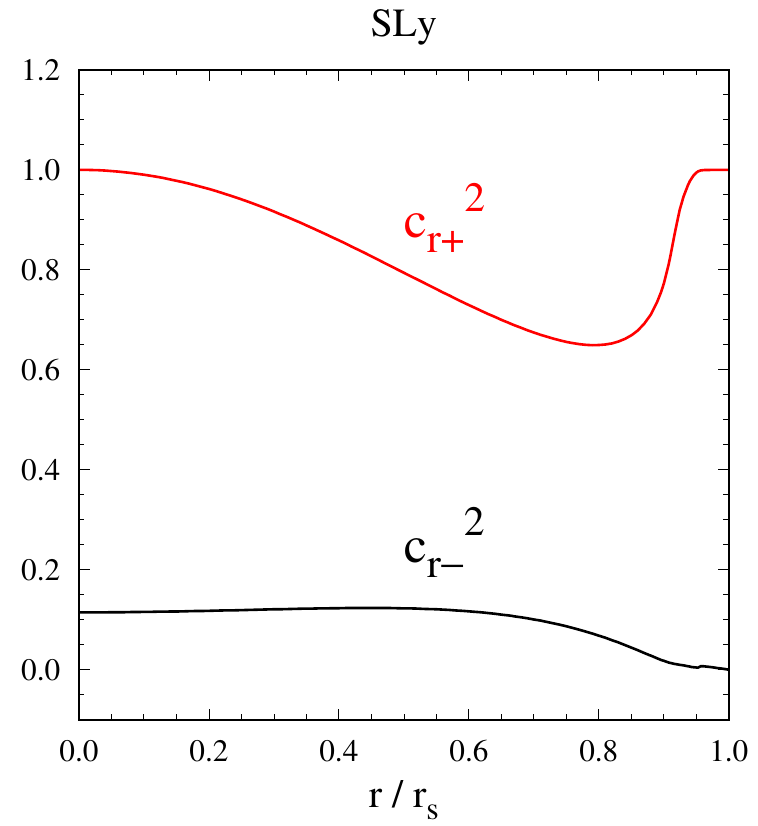}
\includegraphics[height=3.3in,width=3.2in]{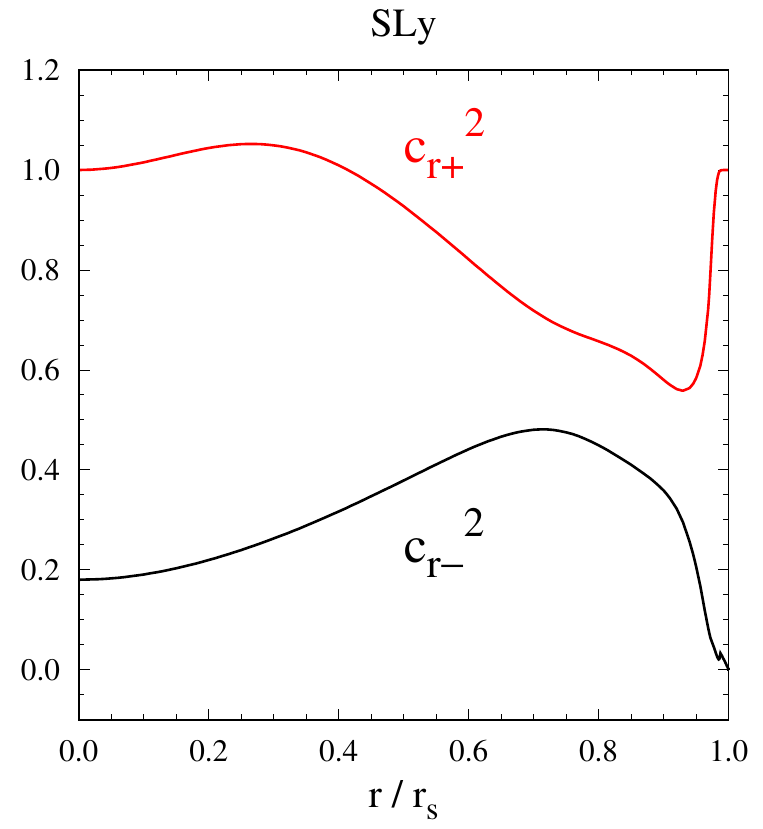}
\end{center}\vspace{-0.5cm}
\caption{\label{fig2}
The radial propagation speed squares $c_{r+}^2$ and 
$c_{r-}^2$ versus $r/r_s$ in NDC theory with 
the branch (\ref{feq}).
We choose the SLy EOS with the central densities
$\rho_c=5\rho_0$ (left) and $\rho_c=15\rho_0$ (right).
}
\end{figure}
%%%%%%%%%%%%%%%%%%%%%%%%%%%%%%%%

In Fig.~\ref{fig2}, we depict $c_{r+}^2$ and $c_{r-}^2$ versus 
$r/r_s$ for the SLy EOS with two different central 
densities $\rho_c$. As estimated from Eq.~(\ref{crp0}), 
$c_{r+}^2$ is close to 1 in the region 
$r/r_s \ll 1$. For $\rho_c=5\rho_0$, $c_{r+}^2$ is in the range 
$0<c_{r+}^2 \le 1$ inside the star and approaches 1 
as $r \to r_s$.
For $\rho_c=15\rho_0$, $c_{r+}^2$ is superluminal 
at small radius, but it becomes subluminal around 
$0.41r_s$ and then reaches the value 1 at $r=r_s$.
The other propagation speed squared $c_{r-}^2$  
is in good agreement with the analytic estimation (\ref{crm0}) around $r=0$. 
As $r$ increases from 0, $c_{r-}^2$ first grows and begins to 
decrease toward the value $+0$ at the surface. 
In Fig.~\ref{fig2}, we can confirm that both $c_{r+}^2$ and $c_{r-}^2$ 
are positive throughout the star interior, so the Laplacian 
instabilities of perturbations $\psi$ and $\delta \phi$ are absent 
along the radial direction.

The propagation speed $c_{\Omega}$ in the angular direction 
is related to the matrix ${\bm M}$ in Eq.~(\ref{Ss2})
as a coefficient containing the term $L$.
This can be obtained by solving 
\be
{\rm det} \left| l^2 f c_{\Omega}^2 {\bm K}
+r^2 {\bm M} \right|=0\,,
\label{detO}
\ee
with the limit $l \to \infty$.
One of the solutions to Eq.~(\ref{detO}) is the matter sound 
speed squared, i.e., 
\be
c_{\Omega 1}^2=c_m^2\,.
\label{com}
\ee
The other two propagation speed squares are given by 
\be
c_{\Omega \pm}^2=\frac{\beta_2 \pm 
\sqrt{\beta_2^2-4\beta_1 \beta_3}}{2\beta_1}\,,
\label{come}
\ee
where
\ba
\beta_1 &=& h(3h-1)\rho \left( 4h-1+16 \pi G_{\rm N} \rho r^2 \right)
-h \left[ \rho+P-h (3\rho+19P) \right]
+Ph \left( 76 h^2-23h+1 \right)
-16P \pi G_{\rm N} hr^2 [ 2\rho+P \nonumber \\
& &-h \left\{ 26 \rho+P(8h+23) \right\} ]
-64P \pi^2 G_{\rm N}^2 hr^4 \left[ 3P(\rho+9P)h^2
-(\rho+P)\{ (3h-1)\rho+(24h-1)P \} \right]\,,\\
\beta_2 &=& \rho \left[ h (3h+1)(5h-3)+12 \pi G_{\rm N} 
(h^2-1) \rho r^2 \right]+Ph \left( 83 h^2+8h-3 \right)
+8 \rho P \pi G_{\rm N}  r^2 [13h^2+6h-3 \nonumber \\
& &+6 \pi G_{\rm N} (h^2-1) \rho r^2]
+4P^2  \pi G_{\rm N} r^2 \left[ 16 h^3+39h^2+12h-3
-12\pi G_{\rm N} \rho r^2 (5h^3-10h^2-5h+2) \right]\nonumber \\
& &-16P^3 \pi^2 G_{\rm N}^2 r^4 
\left( 99 h^3+h^2-15h+3 \right)\,,\\
\beta_3 &=&h \left[ (3\rho+7P)h^2+12Ph-3(\rho+P)  \right]
-16P^2\pi G_{\rm N}h  r^2 [\pi G_{\rm N}r^2 
\{ 3\rho (h^2-1)+P(27h^2+4h+1) \}\nonumber \\
& &
+4h(h-1)]\,.
\ea
On using the background solutions (\ref{fex})-(\ref{Pex}) 
around $r=0$, it follows that 
\ba
c_{\Omega +}^2 &=& 1+{\cal O}(r^2)\,,
\label{cOp}\\
c_{\Omega -}^2 &=& \frac{2P_c}{\rho_c+9P_c}+{\cal O}(r^2)\,,
\label{cOm}
\ea
whose leading-order terms are the same as Eqs.~(\ref{crp0}) 
and (\ref{crm0}), respectively.

Around $r=r_s$, applying the approximation $P/\rho \ll 1$ 
to Eq.~(\ref{come}) gives 
\ba
c_{\Omega +}^2 &=& \frac{h}{h+4\pi G_{\rm N} \rho r^2}
+{\cal O}\left( \frac{P}{\rho} \right)\,,\label{come1}\\
c_{\Omega -}^2 &=& -\frac{3(1-h^2)}{4h (3h-1)}
+{\cal O}\left( \frac{P}{\rho} \right)\,.\label{come2}
\ea
The propagation speed squared (\ref{come1}), which 
corresponds to that of the perturbation $\psi$, is 
subluminal and approaches 1 as $r \to r_s$. 
On the other hand, the leading-order term of Eq.~(\ref{come2}), 
which is associated with the propagation of $\delta \phi$, 
is negative for $1/3<h<1$. 
On the star surface, we have 
\be
c_{\Omega -}^2 (r_s)=-\frac{3{\cal C}(1-{\cal C})}
{2(1-3{\cal C})(1-2{\cal C})}\,,
\label{cOmers}
\ee
where ${\cal C}=G_{\rm N}M_s/r_s$ is the compactness. 
For compact objects satisfying the condition, 
\be
0<{\cal C}<\frac{1}{3}\,,
\ee
it follows that $c_{\Omega -}^2 (r_s)<0$. 
This means that, even though $c_{\Omega -}^2>0$ around $r=0$, 
the sign of $c_{\Omega -}^2$ changes to negative at some radius 
inside the star.

%%%%%%%%%%%%%%%%%%%%%%%%%%%%%%%%
\begin{figure}[h]
\begin{center}
\includegraphics[height=3.3in,width=3.2in]{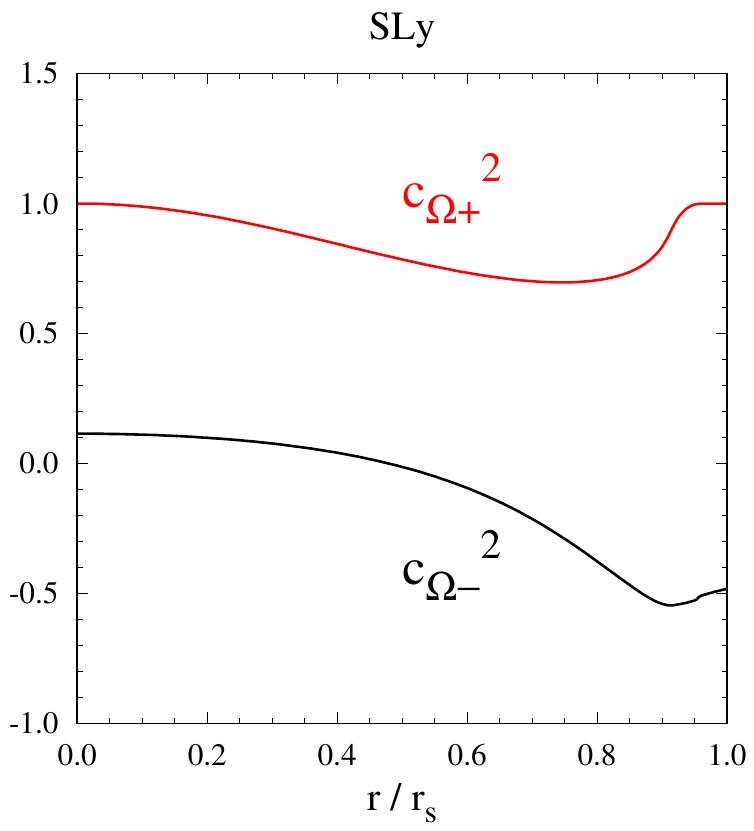}
\includegraphics[height=3.3in,width=3.2in]{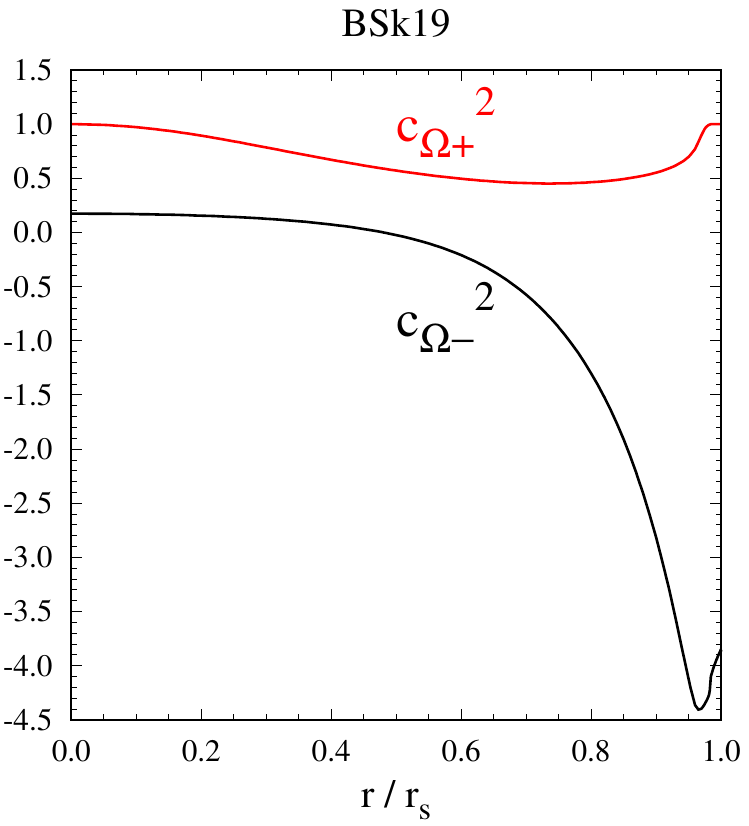}
\end{center}\vspace{-0.5cm}
\caption{\label{fig3} 
(Left)
The angular propagation speed squares $c_{\Omega+}^2$ and 
$c_{\Omega-}^2$ versus $r/r_s$  in NDC theory  
with the branch (\ref{feq}). We choose the SLy EOS with 
$\rho_c=5\rho_0$ (left) and the BSk19 EOS with 
$\rho_c=15\rho_0$ (right).}
\end{figure}
%%%%%%%%%%%%%%%%%%%%%%%%%%%%%%%%

In Fig.~\ref{fig3}, we plot $c_{\Omega+}^2$ and $c_{\Omega-}^2$ versus 
$r/r_s$ for the SLy EOS with $\rho_c=5\rho_0$ (left) and the BSk19 EOS 
with $\rho_c=15\rho_0$ (right). 
For the latter, we use analytic representations of the EOS given in 
Ref.~\cite{Potekhin:2013qqa}.
As estimated from Eqs.~(\ref{cOp}) and (\ref{come1}), $c_{\Omega +}^2$ 
approaches 1 in both limits $r \to 0$ and $r \to r_s$, with 
the stability condition $c_{\Omega +}^2>0$ satisfied inside the star. 
The value of $c_{\Omega-}^2$ around $r=0$ is given by 
$c_{\Omega-}^2 \simeq 2P_c/(\rho_c+9P_c)>0$, but 
it enters the region $c_{\Omega-}^2<0$ at a distance $r_*$.
In the numerical simulation of Fig.~\ref{fig3}, this critical distance 
is $r_*=0.479r_s$ (left) and $r_*=0.481r_s$ (right). 
In Fig.~\ref{fig3}, we observe that $c_{\Omega-}^2$ exhibits some small 
increase around the star surface, but $c_{\Omega-}^2$ is negative
in the region $r_*<r \le r_s$. 
The star compactness in Fig.~\ref{fig3} is given by 
${\cal C}=0.148$ (left) and ${\cal C}=0.276$ (right), in which cases
the analytic estimation (\ref{cOmers}) gives 
$c_{\Omega -}^2(r_s)=-0.483$ (left) and 
$c_{\Omega -}^2(r_s)=-3.89$ (right), respectively. 
They are in good agreement with the numerical values of $c_{\Omega -}^2$ at $r=r_s$.
Hence there are Laplacian instabilities of the field perturbation 
$\delta \phi$ along the angular direction in the region close to the surface of star. 
This means that the background field profile $\phi=\phi(r)$ satisfying the relation 
(\ref{etare}) is unstable against even-parity perturbations for large $l$ modes.

%%%%%%%%%%%%%%%%%%%%%%%%%%%%%%%%
\begin{figure}[h]
\begin{center}
\includegraphics[height=3.2in,width=3.4in]{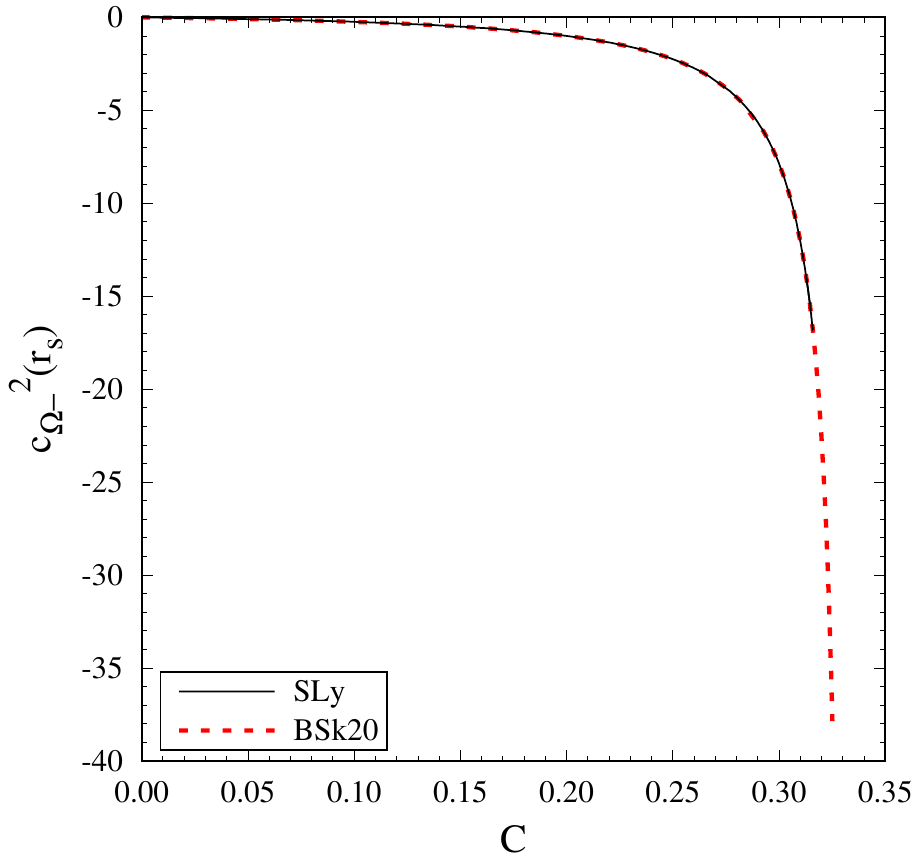}
\includegraphics[height=3.2in,width=3.4in]{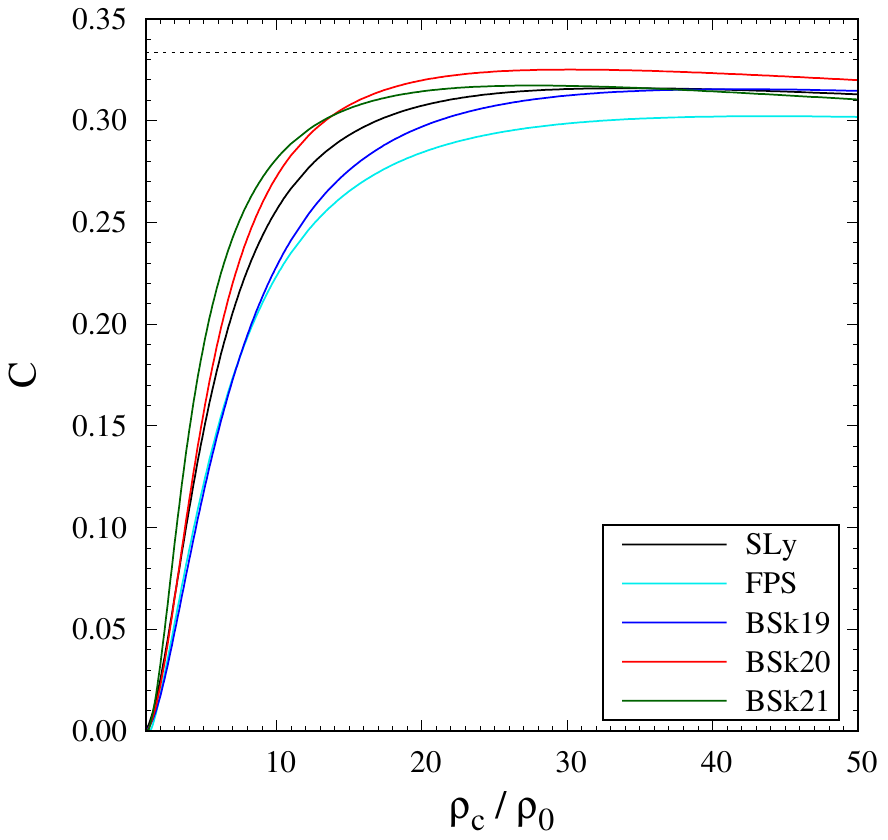}
\end{center}
\caption{\label{fig4} 
(Left)
The angular propagation speed squared $c_{\Omega-}^2$ 
at $r=r_s$ versus the star compactness ${\cal C}$ 
in NDC theory  with the branch (\ref{feq}). 
The solid black and dashed red lines correspond to 
the SLy and BSk20 EOSs, respectively.
(Right)
The compactness ${\cal C}$ versus $\rho_c/\rho_0$ for 
the SLy, FPS, BSk19, BSk20, BSk21 EOSs. 
We also show the border ${\cal C}=1/3$ as a dashed line. 
For the five EOSs with any central density $\rho_c$, ${\cal C}$ 
is smaller than 1/3.}
\end{figure}
%%%%%%%%%%%%%%%%%%%%%%%%%%%%%%%%

In the left panel of Fig.~\ref{fig4}, we depict the numerical values of 
$c_{\Omega -}^2(r_s)$ versus the compactness ${\cal C}$ for the 
SLy and BSk20 EOSs. They are obtained by choosing different central 
densities $\rho_c$ in the range $\rho_0 \le \rho_c \le 50 \rho_0$. 
We confirm that the analytic value of $c_{\Omega -}^2(r_s)$ 
in Eq.~(\ref{cOmers}), which depends on ${\cal C}$ alone, 
is in good agreement with the numerical results. 
For increasing $\rho_c$, ${\cal C}$ first increases with the 
decrease of $c_{\Omega -}^2(r_s)$, but 
there is the saturation for the growth of ${\cal C}$. 
As we observe in the right panel of Fig.~\ref{fig4}, 
the compactness corresponding to the SLy EOS reaches a maximum 
value ${\cal C}_{\rm max}=0.3159$ around $\rho_c=34 \rho_0$.
This situation is similar for other EOSs, with different maximum 
values of ${\cal C}$, e.g., the BSk20 EOS gives ${\cal C}_{\rm max}=0.3251$.
Since ${\cal C}_{\rm max}$ is smaller than $1/3$ for the five EOSs used 
in Fig.~\ref{fig4}, we have $c_{\Omega -}^2(r_s)<0$ for any central density $\rho_c$.
Indeed, the left panel of Fig.~\ref{fig4} shows that, irrespective of the EOSs,  
$c_{\Omega -}^2(r_s)$ is solely determined by the compactness ${\cal C}$ 
and that the Laplacian instability is generally present for ${\cal C}<1/3$. 
This means that not only relativistic but also nonrelativistic stars (${\cal C} \ll 1$) 
in NDC theory are prone to the Laplacian instability along the angular direction.

There may be some specific NS EOSs leading to ${\cal C}$ larger than $1/3$. 
In this case, however, the metric component $h$ is smaller than 
$1/3$ at $r=r_s$. Then, the term ${\cal K}_s$ of Eq.~(\ref{Ks}), which is the dominant 
contribution to ${\cal K}$ in Eq.~(\ref{go3a}) around $r=r_s$, becomes negative 
and hence there is the ghost instability. 
This implies that, even for ${\cal C}>1/3$, the solutions
are plagued by the appearance of ghosts.
Thus, we showed that the hairy compact objects in NDC theory are generally 
subject to instabilities of even-parity perturbations. 

We also make a brief comment on the stability of above hairy solutions 
outside the compact object. Since $P=0$ outside the star, Eq.~(\ref{etare}) 
shows that the background field derivative $\phi'(r)$ vanishes for $r>r_s$.
On using the background Eqs.~(\ref{rheq}) and (\ref{rfeq}),  
all the terms containing $\delta\phi$ and its derivatives in 
Eq.~(\ref{eaction}) disappear for $r>r_s$, so 
the resulting second-order Lagrangian is the same as that in GR with no 
propagation of the scalar field.
Hence the instabilities of hairy solutions are
present only inside the star. 

So far, we have discussed the stability of hairy solutions satisfying Eq.~(\ref{feq}). 
There is also the other branch of Eq.~(\ref{fhre}) characterized by  
\be
\phi'(r)=0\,.
\label{tri}
\ee
For this branch, the metric components $h$ and $f$ 
obey Eqs.~(\ref{rheq}) and (\ref{rfeq}) with $\phi'(r)=\phi''(r)=0$, 
so the background solutions are the same as those in GR 
without the scalar field.
However, we need to caution that the scalar-field perturbation 
$\delta \phi$ still propagates inside the star.
Substituting $\phi'(r)=\phi''(r)=0$ into the second-order 
Lagrangian (\ref{eaction}), we find that the perturbation $\delta \phi$ 
is decoupled from the other fields $\psi$ and $\delta \rho_m$. 
Although the second-order Lagrangian of $\psi$ and $\delta \rho_m$ 
is the same as that in GR, there is the additional scalar-field 
Lagrangian given by
\be
{\cal L}_{\delta \phi}=
e_1 \dot{\delta \phi}^2+e_2 \delta \phi'^2
+\left( e_3+L e_4 \right) \delta \phi^2\,,
\ee
where 
\be
e_1=\frac{4\eta \pi G_{\rm N}\rho\,r^2}{\sqrt{fh}}\,,\qquad
e_2=4\eta \sqrt{fh}\,\pi G_{\rm N} Pr^2\,,\qquad 
e_3=0\,,\qquad
e_4=4\eta \sqrt{\frac{f}{h}}\,\pi G_{\rm N} P\,.
\ee
This shows that, for $\eta \neq 0$, the field perturbation $\delta \phi$ 
propagates inside the star. The propagation speed squares in the 
radial and angular directions are given, respectively, by 
\be
(c_r^2)_{\delta \phi}=-\frac{e_2}{fh e_1}=-\frac{P}{\rho}\,,\qquad 
(c_{\Omega}^2)_{\delta \phi}=-\frac{r^2 e_4}{f e_1}=-\frac{P}{\rho}\,,
\ee
where we have taken the limit $L \to \infty$ for the derivation 
of $(c_{\Omega}^2)_{\delta \phi}$. 
This means that, for a positive perfect-fluid pressure $P$, 
the perturbation $\delta \phi$ is subject to Laplacian instabilities 
inside the star along both radial and angular directions.
Hence the scalar field does not maintain the background 
profile (\ref{tri}). 

%%%%%%%%%%%%%%%%%%%%%%%%%%%%%%%%%%%%%%%%%%
\section{Conclusions}
\label{concludesec}
%%%%%%%%%%%%%%%%%%%%%%%%%%%%%%%%%%%%%%%%%%

We studied the stability of relativistic stars in NDC theory given by 
the action (\ref{NDCaction}) with the coupling (\ref{G5}). 
We dealt with baryonic matter inside the star as a perfect fluid 
described by the action (\ref{SM}).
Besides the trivial branch characterized by $\phi'(r)=0$, 
there is a nontrivial solution endowed with a scalar hair inside compact 
objects. Since the field derivative $\phi'$ is related to the matter pressure $P$ 
according to Eq.~(\ref{etare}), the radius $r_s$ and mass $M_s$ 
of star for the latter branch are determined by integrating 
Eqs.~(\ref{mattereq}), (\ref{feq}), and (\ref{heq}) for a given EOS.
In comparison to GR, the mass-radius relation shifts to 
the region with smaller values of $r_s$ and $M_s$. 

In Sec.~\ref{oddsec}, we first discussed the propagation of 
odd-parity perturbations on the static and spherically-symmetric 
background. For the multipoles $l \geq 2$ the second-order 
action reduces to the form (\ref{Sodd2}), 
with a single dynamical perturbation $\chi$ given by Eq.~(\ref{chiex}).
For the hairy branch in NDC theory, the stability against 
odd-parity perturbations requires that $4\pi G_{\rm N}Pr^2<1$, 
which is well satisfied for NSs with the compactness ${\cal C} \lesssim 1/3$. 
Under this condition the radial propagation speed $c_r$ in the 
odd-parity sector is superluminal, while the angular propagation speed $c_\Omega$ 
is equivalent to that of light.

In Sec.~\ref{evensec}, we analyzed the stability of hairy NS solutions in 
NDC theory against even-parity perturbations. 
In this sector, there are three dynamical degrees of freedom, i.e., 
matter perturbation $\delta \rho_m$, gravitational perturbation $\psi$, 
and scalar-field perturbation $\delta \phi$. 
Under the weak-energy condition $\rho+P>0$, the ghost is absent 
for ${\cal K}>0$, where ${\cal K}$ is given by 
Eq.~(\ref{go3a}). The numerical simulations for several different EOSs 
showed that ${\cal K}$ is positive throughout the star interior. 
For the radial propagation in the even-parity sector, the speed $c_{r1}$ 
associated with $\delta \rho_m$ vanishes by reflecting the fact that  
the matter velocity potential is related to the $\theta, \varphi$ components of 
four velocity. The other two propagation speed squares $c_{r \pm}^2$ 
are both positive, so the Laplacian instabilities are absent for the 
radial propagation of perturbations $\psi$ and $\delta \phi$.

Along the angular direction, the matter perturbation propagates with the sound 
speed squared $c_m^2$ given by Eq.~(\ref{cm}). 
The gravitational perturbation $\psi$ has a positive propagation speed squared 
$c_{\Omega+}^2$.
However, we showed that the speed squared $c_{\Omega-}^2$  
of $\delta \phi$ in the angular direction becomes 
negative around the surface of star with the compactness ${\cal C}$ 
smaller than $1/3$. In particular, the value of $c_{\Omega-}^2$  
at $r=r_s$ is solely determined by the compactness ${\cal C}$, 
see Eq.~(\ref{cOmers}). By choosing five different NS EOSs, we
showed that the Laplacian instability associated with negative  
values of $c_{\Omega-}^2$ is always present, with ${\cal C}<1/3$.
Even for some specific EOS leading to ${\cal C}>1/3$, 
the ghost in the even-parity sector appears around the 
surface of star. We also found that, in the presence of NDCs ($\eta \neq 0$), 
even the branch with $\phi'=0$ is prone to Laplacian instabilities of 
field perturbation $\delta \phi$ inside the star.

These generic instabilities in NDC theory are mostly attributed to the 
nonstandard propagation of $\delta \phi$. 
The propagation is modified by adding a canonical kinetic term $X$ 
in the action (\ref{NDCaction}), but in this case the no-hair theorem of 
Ref.~\cite{Lehebel:2017fag} states that there is only a trivial branch with $\phi'=0$. 
Hence the hairy NS solutions disappear in the presence of the standard kinetic term.
The other possibility for allowing the existence of a nontrivial branch with $\phi' \neq 0$ 
is to add noncanonical kinetic terms like $X^2$ or the cubic Galileon Lagrangian 
$X \square \phi$ to the action (\ref{NDCaction}). 
It will be of interest to study whether stable hairy star solutions exist or not in such 
generalized theories. This is left for a future work. 

%%%%%%%%%%%%%%%%%%
\section*{Acknowledgements}
%%%%%%%%%%%%%%%%%%

RK is supported by the Grant-in-Aid for Young Scientists 
of the JSPS No.\,17K14297 and 20K14471. 
ST is supported by the Grant-in-Aid for Scientific Research Fund of the JSPS No.\,19K03854.

\appendix

\renewcommand{\theequation}{A.\arabic{equation}}
\setcounter{equation}{0}

%%%%%%%%%%%%%%%%%%%%%%%%%
\section*{Appendix: Second-order action of 
even-parity perturbation}
\label{app:coefficients}
%%%%%%%%%%%%%%%%%%%%%%%%%

For NDC theory given by the action (\ref{NDCaction}) with 
(\ref{G5}), the coefficients in the second-order action (\ref{eaction}) are given by 
\ba
& &
a_1=\eta \sqrt{f} h^{3/2} r \phi',\qquad
a_2= \frac{\eta}{2}\sqrt{fh} \left[ 2h r\phi''+(1+h+3rh') \phi' 
\right],\qquad
a_3=-r \sqrt{fh} \left( \frac{1}{16\pi G_{\rm N}}+\frac{3}{4} 
\eta h \phi'^2 \right), \nonumber \\
& &
a_4=\sqrt{fh} \left( \frac{1}{16\pi G_{\rm N}}+\frac{1}{4} 
\eta h \phi'^2 \right)\,,\qquad 
a_5=0,\qquad 
a_6=-\frac{\eta}{4r} \sqrt{\frac{f}{h}} \left[ 2h r\phi''+
\phi' \left( rh'+2h \right) \right], \nonumber\\
& &
a_7=a_3'-\frac{r^2}{2}f_1,\qquad
a_8=-\frac{a_4}{2h},\qquad
a_9=a_4'+\left( \frac{1}{r}-\frac{f'}{2f} \right)a_4,\qquad
a_{10} = \frac{r^2}{2} \sqrt{\frac{f}{h}}, \nonumber\\
& &
b_1=\frac{a_4}{2f},\qquad 
b_2=-\frac{2}{f}a_1,\qquad
b_3=\frac{2}{f} \left( a_1'-a_2 \right),\qquad
b_4=-\frac{2}{f}a_3,\qquad
b_5=-2b_1, \nonumber\\
& &
c_1=-\frac{a_1}{fh},\qquad 
c_2=-\frac{\eta}{2} \sqrt{\frac{h}{f}} \phi' 
\left[ 3hr f'+(3h-1)f \right],\qquad 
c_3=0,\qquad
c_4=\frac{\eta}{4} \sqrt{\frac{h}{f}} \phi' 
\left( f'+\frac{2f}{r} \right),\nonumber\\
& &
c_5=-\left(\frac{f'}{2}+ \frac{f}{r} \right) \sqrt{\frac{h}{f}}
\left( \frac{1}{16\pi G_{\rm N}}+\frac{3}{4} 
\eta h \phi'^2 \right),\qquad 
\tilde{c}_{5}=\frac{r^2 f_1}{\sqrt{f}},
\qquad
c_6=\frac{\sqrt{h}\left\{r f'+f+4\eta \pi G_{\rm N} \phi'^2 [6h(rf'+f)-f]\right\}}
{32\pi G_{\rm N}\sqrt{f}},\nonumber \\
& &
d_1=b_1,\qquad d_2=2hc_4,\qquad 
d_3=-\eta \frac{h^{3/2}\phi'}{\sqrt{f}r^2} \left( rf'+f \right),
\qquad 
d_4=\frac{a_4}{r^2}, \qquad e_1=-\eta \frac{rh'+h-1}{2\sqrt{fh}},\nonumber \\
& &
e_2=\frac{\eta}{2} \sqrt{\frac{h}{f}} \left[ hr f'+(h-1)f \right],\qquad 
e_3=0,\qquad
e_4=\eta \frac{2f(hr f''+fh')-f'(hrf'-frh'-2fh)}{8r f^{3/2} \sqrt{h}},\nonumber \\
& &
f_{1} = -\frac{\rho+P}{2}\sqrt{\frac{f}{h}}\,,\qquad
f_{2}=-\frac{c_m^2 r^2}{2(\rho+P)} \sqrt{\frac{f}{h}}\,,\qquad
f_{3}=-\frac{r^2}{\sqrt{h}}\,.
\ea
Since we did not specify the branches of background solutions, 
these coefficients are valid for both $\phi'(r)=0$ and $\phi'(r) \neq 0$. 

%%%%%%%%%%%%%%%%


\begin{thebibliography}{99}
%%%%%%%%%%%%%%%%

\bibitem{CST}
E.~J.~Copeland, M.~Sami and S.~Tsujikawa,
%``Dynamics of dark energy,''
Int. J. Mod. Phys. D \textbf{15}, 1753-1936 (2006)
%doi:10.1142/S021827180600942X
[arXiv:hep-th/0603057 [hep-th]].

\bibitem{Clifton}
T.~Clifton, P.~G.~Ferreira, A.~Padilla and C.~Skordis,
%``Modified Gravity and Cosmology,''
Phys.\ Rept.\  {\bf 513}, 1 (2012)
[arXiv:1106.2476 [astro-ph.CO]].

\bibitem{Joyce}
A.~Joyce, B.~Jain, J.~Khoury and M.~Trodden,
%``Beyond the Cosmological Standard Model,''
Phys.\ Rept.\  {\bf 568}, 1 (2015)
%doi:10.1016/j.physrep.2014.12.002
[arXiv:1407.0059 [astro-ph.CO]].

\bibitem{Planck}
N.~Aghanim \textit{et al.} [Planck],
%``Planck 2018 results. VI. Cosmological parameters,''
arXiv:1807.06209 [astro-ph.CO].

\bibitem{Abbott2016} 
B.~P.~Abbott {\it et al.} [LIGO Scientific and Virgo Collaborations],
%``Observation of Gravitational Waves from a Binary Black Hole Merger,''
Phys.\ Rev.\ Lett.\  {\bf 116}, 061102 (2016)
%doi:10.1103/PhysRevLett.116.061102
[arXiv:1602.03837 [gr-qc]].
%%CITATION = doi:10.1103/PhysRevLett.116.

\bibitem{GW170817} 
B.~P.~Abbott {\it et al.} [LIGO Scientific and Virgo Collaborations],
%``GW170817: Observation of Gravitational Waves from a Binary Neutron Star Inspiral,''
Phys.\ Rev.\ Lett.\  {\bf 119}, 161101 (2017)
%doi:10.1103/PhysRevLett.119.161101
[arXiv:1710.05832 [gr-qc]].

\bibitem{Berti}
E.~Berti {\it et al.}, 
%``Testing General Relativity with Present and Future Astrophysical Observations,''
Class. Quant. Grav. \textbf{32}, 243001 (2015)
%doi:10.1088/0264-9381/32/24/243001
[arXiv:1501.07274 [gr-qc]].
%670 citations counted in INSPIRE as of 24 Aug 2020

\bibitem{Barack}
L.~Barack {\it et al.}, 
%``Black holes, gravitational waves and fundamental physics: a roadmap,''
Class. Quant. Grav. \textbf{36}, 143001 (2019)
%doi:10.1088/1361-6382/ab0587
[arXiv:1806.05195 [gr-qc]].
%227 citations counted in INSPIRE as of 24 Aug 2020

\bibitem{Israel} 
W.~Israel,
%``Event horizons in static vacuum space-times,''
Phys.\ Rev.\  {\bf 164}, 1776 (1967).

\bibitem{Carter} 
B.~Carter,
%``Axisymmetric Black Hole Has Only 
%Two Degrees of Freedom,''
Phys.\ Rev.\ Lett.\  {\bf 26}, 331 (1971).

\bibitem{Wheeler} 
R.~Ruffini and J.~A.~Wheeler, Phys.\ Today {\bf 24}, 
No. 1, 30 (1971).

\bibitem{Hawking} 
S.~W.~Hawking,
%``Black holes in general relativity,''
Commun.\ Math.\ Phys.\  {\bf 25}, 152 (1972).

\bibitem{Chase} 
J.~E.~Chase, Commun. Math.\ Phys.\ {\bf 19}, 276 (1970).

\bibitem{BekenPRL} 
J.~D.~Bekenstein,
%``Transcendence of the law of baryon-number conservation in black hole physics,''
Phys.\ Rev.\ Lett.\  {\bf 28}, 452 (1972).

\bibitem{Hawking72} 
S.~W.~Hawking,
%``Black holes in the Brans-Dicke theory of gravitation,''
Commun.\ Math.\ Phys.\  {\bf 25}, 167 (1972).
%doi:10.1007/BF01877518

\bibitem{Beken95} 
J.~D.~Bekenstein,
%``Novel ‘‘no-scalar-hair’’ theorem for black holes,''
Phys.\ Rev.\ D {\bf 51}, R6608 (1995).
%doi:10.1103/PhysRevD.51.R6608

\bibitem{Soti12} 
T.~P.~Sotiriou and V.~Faraoni,
%``Black holes in scalar-tensor gravity,''
Phys.\ Rev.\ Lett.\  {\bf 108}, 081103 (2012)
%doi:10.1103/PhysRevLett.108.081103
[arXiv:1109.6324 [gr-qc]].

\bibitem{Rinaldi} 
M.~Rinaldi,
%``Black holes with non-minimal derivative coupling,''
Phys.\ Rev.\ D {\bf 86}, 084048 (2012)
%doi:10.1103/PhysRevD.86.084048
[arXiv:1208.0103 [gr-qc]].

\bibitem{Anabalon} 
A.~Anabalon, A.~Cisterna and J.~Oliva,
%``Asymptotically locally AdS and flat black holes in Horndeski theory,''
Phys.\ Rev.\ D {\bf 89}, 084050 (2014)
%doi:10.1103/PhysRevD.89.084050
[arXiv:1312.3597 [gr-qc]].

\bibitem{Minami13} 
M.~Minamitsuji,
%``Solutions in the scalar-tensor theory with 
%nonminimal derivative coupling,''
Phys.\ Rev.\ D {\bf 89}, 064017 (2014)
%doi:10.1103/PhysRevD.89.064017
[arXiv:1312.3759 [gr-qc]].

\bibitem{Soti1} 
T.~P.~Sotiriou and S.~Y.~Zhou,
%``Black hole hair in generalized scalar-tensor gravity,''
Phys.\ Rev.\ Lett.\  {\bf 112}, 251102 (2014)
%doi:10.1103/PhysRevLett.112.251102
[arXiv:1312.3622 [gr-qc]].

\bibitem{Soti2} 
T.~P.~Sotiriou and S.~Y.~Zhou,
%``Black hole hair in generalized scalar-tensor 
%gravity: An explicit example,''
Phys.\ Rev.\ D {\bf 90}, 124063 (2014)
%doi:10.1103/PhysRevD.90.124063
[arXiv:1408.1698 [gr-qc]].

\bibitem{Babi17} 
E.~Babichev, C.~Charmousis and A.~Leh\'ebel,
%``Asymptotically flat black holes in Horndeski theory and beyond,''
JCAP {\bf 1704}, 027 (2017)
%doi:10.1088/1475-7516/2017/04/027
[arXiv:1702.01938 [gr-qc]].

\bibitem{Horndeski} 
G.~W.~Horndeski,
%``Second-order scalar-tensor field equations in a four-dimensional space,''
Int.\ J.\ Theor.\ Phys.\  {\bf 10}, 363 (1974).

\bibitem{Ho1}
C.~Deffayet, X.~Gao, D.~A.~Steer and G.~Zahariade,
%``From k-essence to generalised Galileons,''
Phys.\ Rev.\ D {\bf 84}, 064039 (2011)
%doi:10.1103/PhysRevD.84.064039
[arXiv:1103.3260 [hep-th]].

\bibitem{Ho2}
T.~Kobayashi, M.~Yamaguchi and J.~'i.~Yokoyama,
%``Generalized G-inflation: Inflation with the most
%general second-order field equations,''
Prog.\ Theor.\ Phys.\  {\bf 126}, 511 (2011)
[arXiv:1105.5723 [hep-th]].

\bibitem{Ho3}
C.~Charmousis, E.~J.~Copeland, A.~Padilla and P.~M.~Saffin,
%``General second order scalar-tensor theory,
%self tuning, and the Fab Four,''
Phys.\ Rev.\ Lett.\  {\bf 108}, 051101 (2012)
[arXiv:1106.2000 [hep-th]].

\bibitem{Hui} 
L.~Hui and A.~Nicolis,
%``No-Hair Theorem for the Galileon,''
Phys.\ Rev.\ Lett.\  {\bf 110}, 241104 (2013)
%doi:10.1103/PhysRevLett.110.241104
[arXiv:1202.1296 [hep-th]].

\bibitem{Babi14} 
E.~Babichev and C.~Charmousis,
%``Dressing a black hole with a time-dependent Galileon,''
JHEP {\bf 1408}, 106 (2014)
%doi:10.1007/JHEP08(2014)106
[arXiv:1312.3204 [gr-qc]].

\bibitem{Koba14} 
T.~Kobayashi and N.~Tanahashi,
%``Exact black hole solutions in shift symmetric scalar\UTF{2013}tensor theories,''
PTEP {\bf 2014}, 073E02 (2014)
%doi:10.1093/ptep/ptu096 
[arXiv:1403.4364 [gr-qc]].

\bibitem{Lefteris} 
C.~Charmousis, T.~Kolyvaris, E.~Papantonopoulos and M.~Tsoukalas,
%``Black Holes in Bi-scalar Extensions of Horndeski Theories,''
JHEP {\bf 1407}, 085 (2014)
%doi:10.1007/JHEP07(2014)085
[arXiv:1404.1024 [gr-qc]].

\bibitem{Babi16} 
E.~Babichev, C.~Charmousis, A.~Leh\'ebel and T.~Moskalets,
%``Black holes in a cubic Galileon universe,''
JCAP {\bf 1609}, 011 (2016)
%doi:10.1088/1475-7516/2016/09/011
[arXiv:1605.07438 [gr-qc]].

\bibitem{Cooney:2009rr} 
A.~Cooney, S.~DeDeo and D.~Psaltis,
%``Neutron Stars in f(R) Gravity with Perturbative Constraints,''
Phys.\ Rev.\ D {\bf 82}, 064033 (2010)
%doi:10.1103/PhysRevD.82.064033
[arXiv:0910.5480 [astro-ph.HE]].

\bibitem{Arapoglu:2010rz} 
A.~S.~Arapoglu, C.~Deliduman and K.~Y.~Eksi,
%``Constraints on Perturbative f(R) Gravity via Neutron Stars,''
JCAP {\bf 1107}, 020 (2011)
%doi:10.1088/1475-7516/2011/07/020
[arXiv:1003.3179 [gr-qc]].

\bibitem{Orellana:2013gn} 
M.~Orellana, F.~Garcia, F.~A.~Teppa Pannia and G.~E.~Romero,
%``Structure of neutron stars in $R$-squared gravity,''
Gen.\ Rel.\ Grav.\  {\bf 45}, 771 (2013)
%doi:10.1007/s10714-013-1501-5
[arXiv:1301.5189 [astro-ph.CO]].

\bibitem{Astashenok:2013vza} 
A.~V.~Astashenok, S.~Capozziello and S.~D.~Odintsov,
%``Further stable neutron star models from f(R) gravity,''
JCAP {\bf 1312}, 040 (2013)
%doi:10.1088/1475-7516/2013/12/040
[arXiv:1309.1978 [gr-qc]].

\bibitem{Yazadjiev:2014cza} 
S.~S.~Yazadjiev, D.~D.~Doneva, K.~D.~Kokkotas and K.~V.~Staykov,
%``Non-perturbative and self-consistent models of neutron 
%stars in R-squared gravity,''
JCAP {\bf 1406}, 003 (2014)
%doi:10.1088/1475-7516/2014/06/003
[arXiv:1402.4469 [gr-qc]].

\bibitem{Resco:2016upv} 
M.~Aparicio Resco, A.~de la Cruz-Dombriz, F.~J.~Llanes Estrada 
and V.~Zapatero Castrillo,
%``On neutron stars in $f(R)$ theories: Small radii, large masses 
%and large energy emitted in a merger,''
Phys.\ Dark Univ.\  {\bf 13}, 147 (2016)
%doi:10.1016/j.dark.2016.07.001
[arXiv:1602.03880 [gr-qc]].

\bibitem{Kase:2019dqc}
R.~Kase and S.~Tsujikawa,
%``Neutron stars in $f(R)$ gravity and scalar-tensor theories,''
JCAP \textbf{09}, 054 (2019)
%doi:10.1088/1475-7516/2019/09/054
[arXiv:1906.08954 [gr-qc]].

\bibitem{Dohi:2020bfs}
A.~Dohi, R.~Kase, R.~Kimura, K.~Yamamoto and M.~a.~Hashimoto,
%``Neutron star cooling in modified gravity theories,''
arXiv:2003.12571 [gr-qc].

\bibitem{Damour} 
T.~Damour and G.~Esposito-Farese,
%``Nonperturbative strong field effects 
%in tensor - scalar theories of gravitation,''
Phys.\ Rev.\ Lett.\  {\bf 70}, 2220 (1993).

\bibitem{Damour2} 
T.~Damour and G.~Esposito-Farese,
%``Tensor - scalar gravity and binary pulsar experiments,''
Phys.\ Rev.\ D {\bf 54}, 1474 (1996)
[gr-qc/9602056].

\bibitem{Sotani:2012eb}
H.~Sotani,
%``Slowly Rotating Relativistic Stars in Scalar-Tensor Gravity,''
Phys. Rev. D \textbf{86}, 124036 (2012)
%doi:10.1103/PhysRevD.86.124036
[arXiv:1211.6986 [astro-ph.HE]].

\bibitem{Doneva:2013qva}
D.~D.~Doneva, S.~S.~Yazadjiev, N.~Stergioulas and K.~D.~Kokkotas,
%``Rapidly rotating neutron stars in scalar-tensor theories of gravity,''
Phys. Rev. D \textbf{88}, 084060 (2013)
%doi:10.1103/PhysRevD.88.084060
[arXiv:1309.0605 [gr-qc]].

\bibitem{Pani:2014jra}
P.~Pani and E.~Berti,
%``Slowly rotating neutron stars in scalar-tensor theories,''
Phys. Rev. D \textbf{90}, 024025 (2014)
%doi:10.1103/PhysRevD.90.024025
[arXiv:1405.4547 [gr-qc]].

\bibitem{Harada:1998ge} 
T.~Harada,
%``Neutron stars in scalar tensor theories of gravity 
%and catastrophe theory,''
Phys.\ Rev.\ D {\bf 57}, 4802 (1998)
%doi:10.1103/PhysRevD.57.4802
[gr-qc/9801049].

\bibitem{Novak:1998rk} 
J.~Novak,
%``Neutron star transition to strong scalar field state 
%in tensor scalar gravity,''
Phys.\ Rev.\ D {\bf 58}, 064019 (1998)
% doi:10.1103/PhysRevD.58.064019
[gr-qc/9806022].
%%CITATION = doi:10.1103/PhysRevD.58.064019;%%

\bibitem{Silva:2014fca} 
H.~O.~Silva, C.~F.~B.~Macedo, E.~Berti and L.~C.~B.~Crispino,
%``Slowly rotating anisotropic neutron stars in general relativity 
%and scalar\UTF{2013}tensor theory,''
Class.\ Quant.\ Grav.\  {\bf 32}, 145008 (2015)
% doi:10.1088/0264-9381/32/14/145008
[arXiv:1411.6286 [gr-qc]].

\bibitem{Freire:2012mg} 
P.~C.~C.~Freire {\it et al.},
%``The relativistic pulsar-white dwarf binary PSR J1738+0333 II. 
%The most stringent test of scalar-tensor gravity,''
Mon.\ Not.\ Roy.\ Astron.\ Soc.\  {\bf 423}, 3328 (2012)
%doi:10.1111/j.1365-2966.2012.21253.x
[arXiv:1205.1450 [astro-ph.GA]].

\bibitem{Kase:2020qvz}
R.~Kase, R.~Kimura, S.~Sato and S.~Tsujikawa,
%``Stability of relativistic stars with scalar hairs,''
arXiv:2007.09864 [gr-qc] 
({\tt Physical Review D to appear}).

\bibitem{Lehebel:2017fag}
A.~Leh\'ebel, E.~Babichev and C.~Charmousis,
%``A no-hair theorem for stars in Horndeski theories,''
JCAP \textbf{07}, 037 (2017)
%doi:10.1088/1475-7516/2017/07/037
[arXiv:1706.04989 [gr-qc]].

\bibitem{Cisterna:2015yla}
A.~Cisterna, T.~Delsate and M.~Rinaldi,
%``Neutron stars in general second order scalar-tensor theory: 
%The case of nonminimal derivative coupling,''
Phys. Rev. D \textbf{92}, 044050 (2015)
%doi:10.1103/PhysRevD.92.044050
[arXiv:1504.05189 [gr-qc]].

\bibitem{Cisterna:2016vdx}
A.~Cisterna, T.~Delsate, L.~Ducobu and M.~Rinaldi,
%``Slowly rotating neutron stars in the nonminimal 
%derivative coupling sector of Horndeski gravity,''
Phys. Rev. D \textbf{93}, 084046 (2016)
%doi:10.1103/PhysRevD.93.084046
[arXiv:1602.06939 [gr-qc]].

\bibitem{Maselli:2016gxk}
A.~Maselli, H.~O.~Silva, M.~Minamitsuji and E.~Berti,
%``Neutron stars in Horndeski gravity,''
Phys. Rev. D \textbf{93}, 124056 (2016)
%doi:10.1103/PhysRevD.93.124056
[arXiv:1603.04876 [gr-qc]].

\bibitem{Blazquez}
J.~L.~Blazquez-Salcedo and K.~Eickhoff,
%``Axial quasinormal modes of static neutron stars in the nonminimal 
%derivative coupling sector of Horndeski gravity: 
%spectrum and universal relations for realistic equations of state,''
Phys. Rev. D \textbf{97}, 104002 (2018)
%doi:10.1103/PhysRevD.97.104002
[arXiv:1803.01655 [gr-qc]].

\bibitem{Ogawa}
H.~Ogawa, T.~Kobayashi and T.~Suyama,
%``Instability of hairy black holes in shift-symmetric Horndeski theories,''
Phys. Rev. D \textbf{93}, 064078 (2016)
%doi:10.1103/PhysRevD.93.064078
[arXiv:1510.07400 [gr-qc]].

\bibitem{Takahashi}
K.~Takahashi and T.~Suyama,
%``Linear perturbation analysis of hairy black holes in shift-symmetric 
%Horndeski theories: Odd-parity perturbations,''
Phys. Rev. D \textbf{95}, 024034 (2017)
%doi:10.1103/PhysRevD.95.024034
[arXiv:1610.00432 [gr-qc]].

\bibitem{Sorkin}
B.~F.~Schutz and R.~Sorkin,
%``Variational aspects of relativistic field theories,
%with application to perfect fluids,''
Annals Phys.\  {\bf 107}, 1 (1977).

\bibitem{Brown} 
J.~D.~Brown,
%``Action functionals for relativistic perfect fluids,''
Class.\ Quant.\ Grav.\  {\bf 10}, 1579 (1993)
%doi:10.1088/0264-9381/10/8/017
[gr-qc/9304026].

\bibitem{DGS}
A.~De Felice, J.~M.~Gerard and T.~Suyama,
%``Cosmological perturbations of a perfect fluid and 
%noncommutative variables,''
Phys.\ Rev.\ D {\bf 81}, 063527 (2010)
%doi:10.1103/PhysRevD.81.063527
[arXiv:0908.3439 [gr-qc]].

\bibitem{Amendola:2020ldb}
L.~Amendola and S.~Tsujikawa,
%``Scaling solutions and weak gravity in dark energy with 
%energy and momentum couplings,''
JCAP \textbf{06}, 020 (2020)
%doi:10.1088/1475-7516/2020/06/020
[arXiv:2003.02686 [gr-qc]].

\bibitem{Haensel:2004nu}
P.~Haensel and A.~Y.~Potekhin,
%``Analytical representations of unified equations of state of neutron-star matter,''
Astron. Astrophys. \textbf{428}, 191-197 (2004)
%doi:10.1051/0004-6361:20041722
[arXiv:astro-ph/0408324 [astro-ph]].

\bibitem{Regge:1957td} 
T.~Regge and J.~A.~Wheeler,
%``Stability of a Schwarzschild singularity,''
Phys.\ Rev.\  {\bf 108}, 1063 (1957).
%%CITATION = doi:10.1103/PhysRev.108.1063;%%
%1200 citations counted in INSPIRE as of 14 Dec 2017

\bibitem{Zerilli:1970se} 
F.~J.~Zerilli,
%``Effective potential for even parity Regge-Wheeler 
%gravitational perturbation equations,''
Phys.\ Rev.\ Lett.\  {\bf 24}, 737 (1970).
%%CITATION = doi:10.1103/PhysRevLett.24.737;%%
%495 citations counted in INSPIRE as of 14 Dec 2017

\bibitem{Moncrief}
V.~Moncrief,
%``Gravitational perturbations of spherically symmetric systems. I. The exterior problem.,''
Annals Phys. \textbf{88}, 323-342 (1974).
%doi:10.1016/0003-4916(74)90173-0

\bibitem{Kobayashi:2012kh}
T.~Kobayashi, H.~Motohashi and T.~Suyama,
%``Black hole perturbation in the most general scalar-tensor theory 
%with second-order field equations I: the odd-parity sector,''
Phys. Rev. D \textbf{85}, 084025 (2012)
%doi:10.1103/PhysRevD.85.084025
[arXiv:1202.4893 [gr-qc]].

\bibitem{DeFelice:2011ka}
A.~De Felice, T.~Suyama and T.~Tanaka,
%``Stability of Schwarzschild-like solutions in f(R,G) gravity models,''
Phys. Rev. D \textbf{83}, 104035 (2011)
%doi:10.1103/PhysRevD.83.104035
[arXiv:1102.1521 [gr-qc]].

\bibitem{Kase:2014baa}
R.~Kase, L.~A.~Gergely and S.~Tsujikawa,
%``Effective field theory of modified gravity on the spherically symmetric 
%background: leading order dynamics and the odd-type perturbations,''
Phys. Rev. D \textbf{90}, 124019 (2014)
%doi:10.1103/PhysRevD.90.124019
[arXiv:1406.2402 [hep-th]].

\bibitem{Kase:2018aps}
R.~Kase and S.~Tsujikawa,
%``Dark energy in Horndeski theories after GW170817: A review,''
Int. J. Mod. Phys. D \textbf{28}, 1942005 (2019).
%doi:10.1142/S0218271819420057
[arXiv:1809.08735 [gr-qc]].

\bibitem{Kobayashi:2014wsa}
T.~Kobayashi, H.~Motohashi and T.~Suyama,
%``Black hole perturbation in the most general scalar-tensor theory with 
%second-order field equations II: the even-parity sector,''
Phys. Rev. D \textbf{89}, 084042 (2014)
%doi:10.1103/PhysRevD.89.084042
[arXiv:1402.6740 [gr-qc]].

\bibitem{Potekhin:2013qqa}
A.~Y.~Potekhin, A.~F.~Fantina, N.~Chamel, J.~M.~Pearson and S.~Goriely,
%``Analytical representations of unified equations of state for neutron-star matter,''
Astron. Astrophys. \textbf{560}, A48 (2013)
%doi:10.1051/0004-6361/201321697
[arXiv:1310.0049 [astro-ph.SR]].

\end{thebibliography}
\end{document}